\documentclass[12pt]{article}
\pdfoutput=1

\usepackage{amsmath,amssymb,amscd}
\usepackage{listings}
\usepackage{caption}
\usepackage{dsfont}
\usepackage{slashed}
\usepackage{color}
\usepackage{ulem}

\usepackage[pdftex]{graphicx}
\usepackage{epstopdf}
\usepackage{subfigure}
\usepackage{epsfig}
\usepackage{listings}
\usepackage{caption}
\usepackage{cite}

\usepackage{multirow}

\newcommand{\beq}{\begin{equation}}
\newcommand{\eeq}{\end{equation}}
\newcommand{\nbea}{\begin{align*}}
\newcommand{\neea}{\end{align*}}
\newcommand{\nbeq}{\begin{equation*}}
\newcommand{\neeq}{\end{equation*}}

 \usepackage{multirow}
\usepackage{array}
\newcolumntype{M}[1]{>{\centering\arraybackslash}m{#1}}
\newcolumntype{N}{@{}m{0pt}@{}}

\numberwithin{equation}{section}

\begin{document}


\pagestyle{empty}

\baselineskip=21pt
\rightline{\footnotesize KCL-PH-TH/2022-02, CERN-TH-2022-011}
\vskip 0.25in

\begin{center}

{\large {\bf Light-by-Light Scattering at Future $e^+e^-$ Colliders}}

\vskip 0.2in

 {\bf John~Ellis}~$^{1,2,3}$,~
   {\bf Nick~E.~Mavromatos}~$^{1,4}$,
 {\bf Philipp~Roloff}~$^5$
and {\bf Tevong~You}~$^{3,6}$

\vskip 0.1in

{\small {\it

$^1${Theoretical Particle Physics and Cosmology Group, Physics Department, \\
King's College London, London WC2R 2LS, UK}\\
\vspace{0.25cm}
$^2${  National Institute of Chemical Physics and Biophysics, R{\" a}vala 10, 10143 Tallinn, Estonia} \\
\vspace{0.25cm}
$^3${Theoretical Physics Department, CERN, CH-1211 Geneva 23, Switzerland}\\
\vspace{0.25cm}
$^4${Physics Department, School of Applied Mathematical and Natural Sciences, 
National Technical University of Athens, Zografou Campus, GR 157 80, Athens, Greece}\\
\vspace{0.25cm}
$^5${Experimental Physics Department, CERN, CH-1211 Geneva 23, Switzerland} \\
\vspace{0.25cm}
$^6${DAMTP, University of Cambridge, Wilberforce Road, Cambridge, CB3 0WA, UK; \\
Cavendish Laboratory, University of Cambridge, J.J. Thomson Avenue, \\ 
\vspace{-0.25cm}
Cambridge, CB3 0HE, UK}
}}

\vskip 0.2in

{\bf Abstract}
\end{center}

We study the sensitivity of possible CLIC and FCC-ee measurements of light-by-light scattering to old and new physics,
including the Heisenberg-Euler Lagrangian in the Standard Model with possible contributions from loops of
additional charged particles or magnetic monopoles, the Born-Infeld extension of QED, and
effective dimension-8 operators involving four electromagnetic field strengths as could appear in the
Standard Model Effective Field Theory. We find that FCC-ee measurements at 365~GeV and CLIC measurements at
350~GeV would be sensitive to new physics scales of half a TeV in the dimension-8 operator coefficients, and that CLIC
measurements at 1.4 TeV or 3~TeV would be sensitive to new physics scales $\sim 2$ TeV or 5 TeV at 95\% CL, 
corresponding to probing loops of new particles with masses up to $\sim 3.7$ TeV for large charges and/or multiple
species. Within Born-Infeld theory,
the $95\%$ CL sensitivities would range from $\sim 300$~GeV to 1.3 or 2.8~TeV for the high-energy CLIC options. Measurements
of light-by-light scattering would not exclude monopole production at FCC-hh, except in the context of
Born-Infeld theory.




\vskip 0.25in

\leftline{ {
March 2022}}

\newpage
\pagestyle{plain}

\section{Introduction}

Light-by-light scattering was first discussed theoretically in the 1930s, from two different points of view.
On the one hand, Heisenberg and Euler~\cite{HeisenbergEuler36} considered how the quantum effects of electron loop diagrams
would induce $\gamma \gamma \to \gamma \gamma$ scattering and higher-order interactions.
On the other hand, Born and Infeld~\cite{BornInfeld34} proposed a nonlinear extension of QED, motivated by the
`unitarian' idea that there should be a maximum electric field analogous to the maximum velocity
provided by the speed of light. Born-Infeld theory predicted characteristic higher-order interactions
that could be probed in light-by-light scattering. 

There has been much subsequent theoretical work on both these approaches to light-by-light
scattering. For example,  
the quantum effects of heavier charged particles in the Standard Model
such as $W^\pm$ bosons and top quarks have been calculated, as well as the leading QCD corrections to quark
loops~\cite{Bernetal}. In the context of supergravity theories, light-by-light scattering has been argued~\cite{Duff}  to provide 
interesting connections between self-duality, helicity and supersymmetry.
Moreover, it was discovered that the Born-Infeld nonlinear modification of QED emerges
naturally in string~\cite{FT85} and brane models~\cite{tseytlin}. In the latter case, the maximum electric field is associated
with the fact that branes have a maximum velocity equal to that of light~\cite{Bachas95}, a stunning vindication of
the `unitarian' intuition of Born and Infeld~\cite{BornInfeld34}. More generally, dispersion relations and crossing symmetry have recently been used to derive new sum rules and positivity bounds on higher-dimensional operators in light-by-light scattering amplitudes~\cite{Henriksson:2021ymi}.   

Over the years, there have also been several experimental efforts to probe possible nonlinear effects
in electrodynamics~\cite{Marklund}. Studies have been made of possible effects on the spectra of electronic and
muonic atoms~\cite{RSG} (see, however,~\cite{CK}), as well as photon splitting in atomic physics~\cite{split}, 
birefringence effects~\cite{PVLAS14}, vacuum dichroism
and measurements of the Lamb shift~\cite{Fouche16}. These constrained various possible nonlinear effects in
electrodynamics. However, the tightest constraint on the energy scale in the Born-Infeld Lagrangian
was $\gtrsim 100$~MeV from Lamb shift measurements~\cite{Fouche16}, orders of magnitude below the scale at 
which string effects could possibly appear.

Following a suggestion by d'Enterria and Silveira~\cite{dES13}, the ATLAS Collaboration made a
first measurement of light-by-light scattering in heavy-ion collisions at the LHC~\cite{ATLAS17}, which has been followed by a
measurement by the CMS Collaboration~\cite{CMS18}. Their results are in
good agreement with the Heisenberg-Euler prediction based on loops of Standard Model particles,
and can be used to constrain possible nonlinear extensions of electrodynamics~\cite{EMY2}. In particular,
it was shown in~\cite{EMY2} that the ATLAS data imposed a lower limit $\gtrsim 100$~GeV on the energy
scale in the Born-Infeld theory, much closer to the threshold where string effects might conceivably appear. More recently, 
the TOTEM and CMS Collaborations have published the first results from a search for light-by-light scattering in 
proton-proton collisions at the LHC~\cite{TOTEM:2021kin}, which can also be used to constrain possible extensions of the Standard Model,
as discussed below.

The LHC constraints are particularly interesting in the context of Born-Infeld extensions of the Standard Model,
which possess finite-energy monopole solutions~\cite{ArunasalamKobakhidze17} of Cho-Maison type~\cite{cho-maison}~\footnote{For a discussion on other types of low-scale magnetic monopole solutions that could exist in extensions of the Standard Model the reader is referred to the  recent review~\cite{mavmit}, and references therein.}.
The mass of such a monopole gets a significant contribution from the corresponding 
Born-Infeld parameter~\cite{ArunasalamKobakhidze17,EMY2}, and the lower limit on the Born-Infeld scale from 
ATLAS data implies~\cite{EMY2} that monopole solutions of the
Born-Infeld extension of the Standard Model should have masses $\gtrsim 11$~\cite{ArunasalamKobakhidze17} to 14~TeV~\cite{sarkar}, 
rendering their production at  LHC impossible~\cite{atlasmono,MoEDAL}.\footnote{See the discussion in Section~\ref{sec:magmon} below.}
However, they could still be within reach of the Future Circular Collider proton-proton option (FCC-hh)~\cite{FCC-hh}, 
if the Born-Infeld scale is within a factor of a few of the ATLAS limit~\cite{EMY2}.

We discuss here the prospective sensitivities of proposed $e^+ e^-$ colliders to light-by-light scattering
and their corresponding sensitivities to physics beyond the Standard Model, such as Born-Infeld theory and
general forms of dimension-8 interactions in the Standard Model Effective Field Theory (SMEFT)~\footnote{Light-by-light scattering may also be used to constrain axion-like particles, see e.g.~\cite{Knapen:2016moh}.}.
CLIC is a proposed $e^+ e^-$ linear collider designed to achieve a centre-of mass energy of 3~TeV~\cite{CLIC},
whereas FCC-ee is a circular $e^+ e^-$ collider capable of reaching 365~GeV in the centre of mass~\cite{FCC:2018evy}.
Such high-energy $e^+ e^-$ colliders generate $\gamma \gamma$ collisions with luminosities that can be calculated accurately
via the equivalent photon approximation (EPA) and Beamstrahlung. Moreover, the
effects of dimension-8 operators such as those in Born-Infeld theory grow rapidly with
energy, increasing the sensitivities of measurements at high energies.  CLIC combines
these two advantages and could therefore be expected to have the greatest sensitivity to nonlinear electrodynamics of any 
accelerator currently proposed, as we explore in this paper.

In a spirit of generality, in addition to the Heisenberg-Euler Lagrangian in the Standard Model,
we consider the possible sensitivity of CLIC and FCC-ee to the coefficients of arbitrary combinations of the
two independent parity-conserving dimension-8 operators that could mediate 
$\gamma \gamma \to \gamma \gamma$ scattering:
\begin{equation}
\mathcal{L}_\text{EFT} \; \supset \; c_1 F_{\mu\nu}F^{\mu\nu} F_{\rho\sigma} F^{\rho\sigma} + c_2 F_{\mu\nu} F^{\nu\rho} F_{\rho\lambda} F^{\lambda\mu} \, .
\label{generaldim8}
\end{equation}
The coefficients $c_{1,2}$ would receive contributions from any new particles beyond the Standard Model,
related directly to their masses and electric charges - see (\ref{cCHE}, \ref{ciHE}) - and can also be related
directly to the scale parameter in a Born-Infeld Lagrangian - see (\ref{cibeta}) - and to the masses and $\gamma \gamma$
couplings of heavy bosons as well as to the possible effects of magnetic monopoles. 

As we show later, the CLIC sensitivities to the coefficients $c_{1,2}$ extend to energy scales of nonlinearity
that are of significant potential interest to string and brane scenarios. In particular, CLIC operating at 3 (1.4) TeV
would be sensitive at the 95\% CL to a Born-Infeld scale of 2.8 (1.3) TeV, showing that it could cover
the range where a monopole in a Born-Infeld extension of the Standard Model could be discovered at FCC-hh. On the other hand,
the sensitivities of CLIC operating at 350~GeV or FCC-ee operating at 365~GeV would be limited to 
Born-Infeld scales $\sim 300$~GeV {that are already excluded by~\cite{TOTEM:2021kin}}.

The layout of this paper is as follows. In Section~2 we discuss the theories behind several possible contributions to
light-by-light scattering, including the SM, dimension-8 operators, the Heisenberg-Euler loop contributions, Born-Infeld
theory and loops of magnetic monopoles. Then, in Section~3 we discuss the possible measurements of $\gamma \gamma \to \gamma \gamma$
at CLIC and FCC-ee, and in Section~4 we discuss the sensitivities to new physics that they would offer. Finally, Section~5
summarizes our conclusions.

\section{Contributions to Light-by-Light Scattering}

\subsection{The Standard Model}
\label{subsec:SM}

\begin{figure}[h!]
\begin{center}
\includegraphics[scale=0.4]{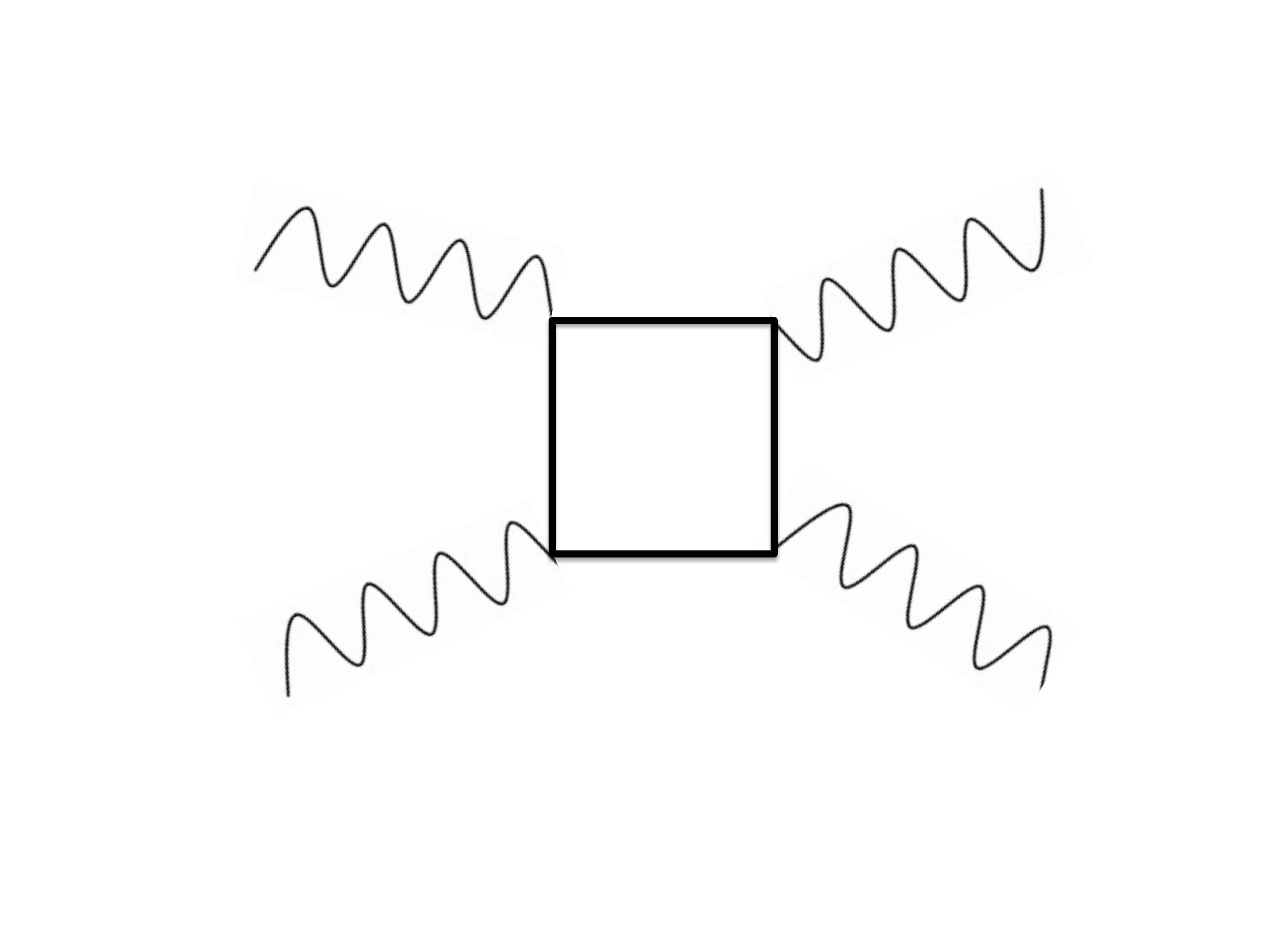}
\vspace{-1.5cm}
\caption{\it Contribution to light-by-light scattering: $\gamma \, \gamma \rightarrow \gamma\, \gamma$ induced by
a fermion loop.}
\label{fig:ggloop}
\end{center}
\end{figure}

There are unavoidable Heisenberg-Euler
contributions to $\gamma \gamma \to \gamma \gamma$ scattering from loops of Standard
Model particles, as illustrated in Fig.~\ref{fig:ggloop} for the case of a fermion loop.

\begin{figure}[h!]
\begin{center}
\includegraphics[scale=0.35]{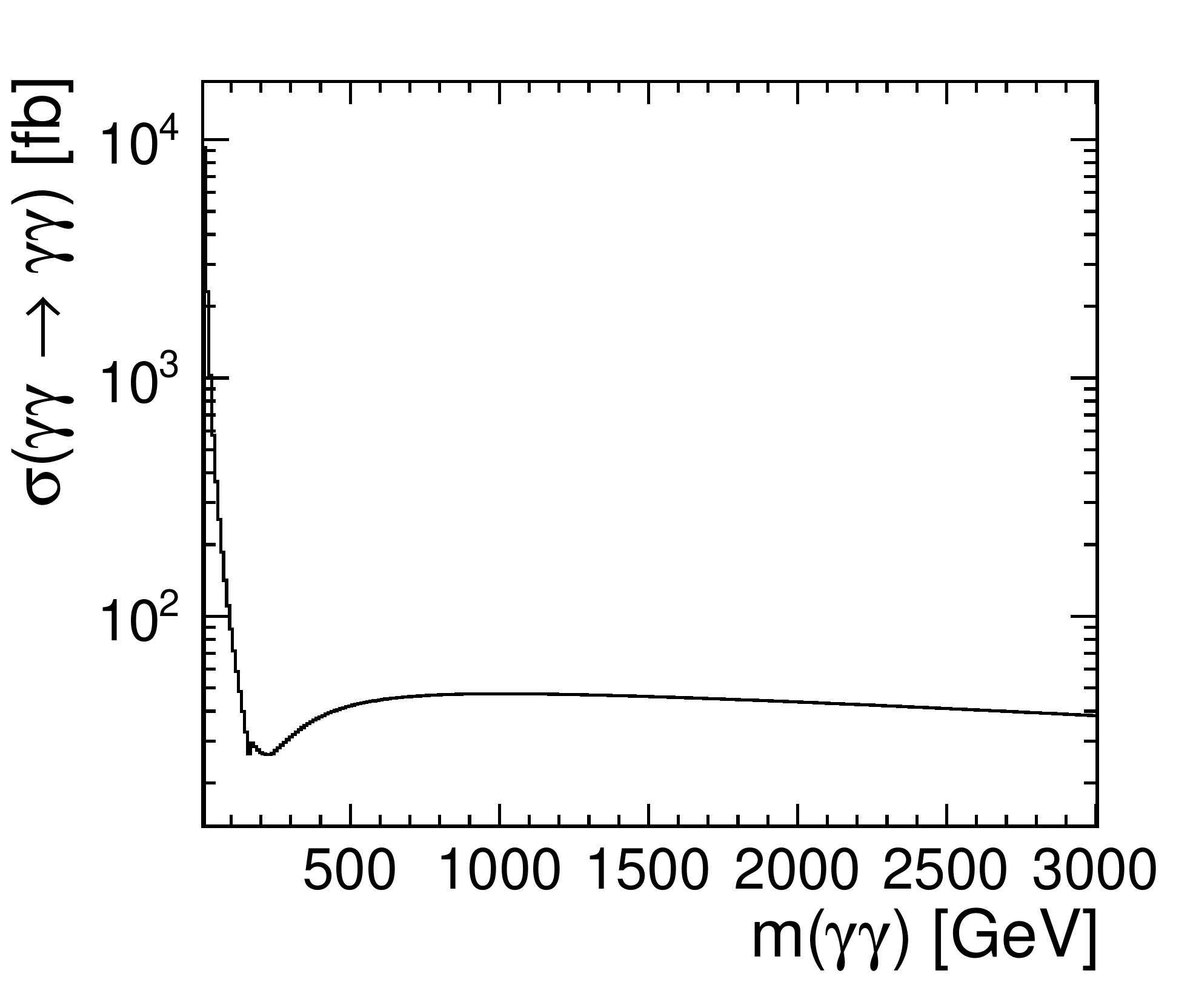}
\includegraphics[scale=0.35]{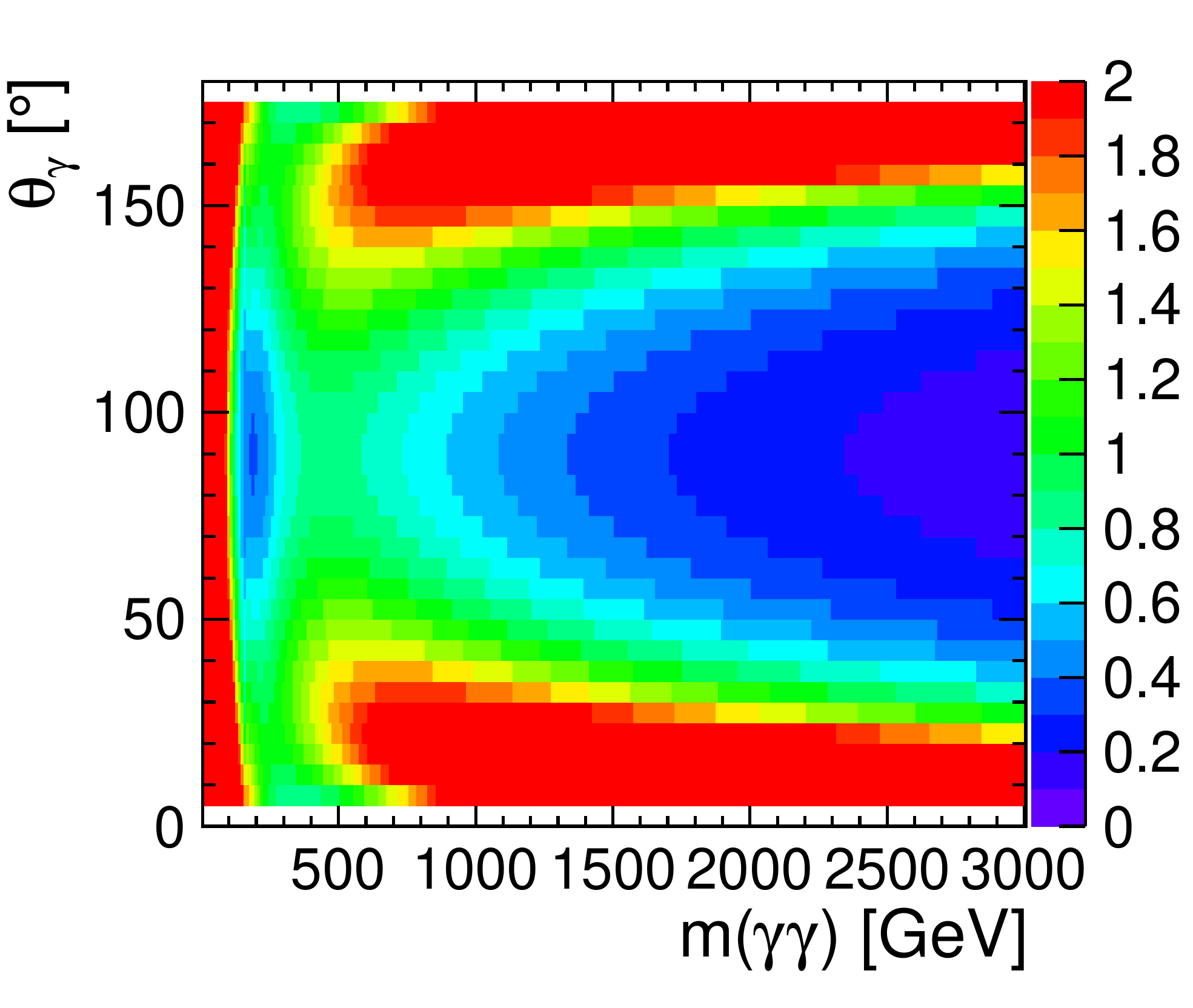}
\caption{\it Left panel: Dependence on the centre-of-mass energy $\sqrt{s}$
of the Heisenberg-Euler cross-section calculated in the photon polar angle region $5^o < \theta_\gamma < 175^o$, showing the effects of the $W^+ W^-$ and ${\bar t} t$
intermediate states. Right panel: The angular distribution in the $\gamma \gamma$ centre of mass as a function of $\sqrt{s}$,
displaying forward-backward peaking except in the neighbourhoods of the $W^+ W^-$ and ${\bar t} t$ thresholds. The colours and numbers represent the cross-section in femtobarns in the corresponding angular bin.
Both panels were generated using the {\tt SANC} code~\protect\cite{SANC}.
}
\label{fig:sanc}
\end{center}
\end{figure}

We calculate the loops of fermions and vector bosons in the Standard Model using the {\tt SANC} code~\cite{SANC}. The left panel of Fig.~\ref{fig:sanc} shows as a solid line the dependence on $m_{\gamma \gamma} = \sqrt{\hat s}$ of the $\gamma \gamma \to \gamma \gamma$ scattering cross-section in the Standard Model. Following a steep fall at lower
$m_{\gamma \gamma}$ where the cross-section is dominated by light-fermion loops, we see a small glitch
around $m_{\gamma \gamma} = 2 m_W \simeq 160$~GeV, the $W^+ W^-$ threshold. This is followed by a prominent rise in the
cross-section above the $t {\bar t}$ threshold at $m_{\gamma \gamma} \gtrsim 350$~GeV.

The right panel of Fig.~\ref{fig:sanc} shows the angular distribution in the centre of mass of $\gamma \gamma \to \gamma \gamma$ scattering in the Standard Model as a function of $m_{\gamma \gamma}$. The colours
and numbers represent the cross-section in femtobarns in the corresponding angular bin. We see that
there is strong forward-backward peaking except in the regions of the $W^+ W^-$ and ${\bar t} t$ thresholds. A similar feature would
appear close to any threshold for new charged particles.

\subsection{Dimension-8 Operator Contributions}

The contributions of particles with masses $\gg m_{\gamma \gamma}$ may be parametrized in terms of the
dimension-8 Lagrangian terms (\ref{generaldim8}). The angular cross-section for unpolarised light-by-light scattering obtained from
these Lagrangian terms may be written as
\begin{equation}
\frac{d\sigma}{d\Omega} = \frac{1}{16\pi^2 {\hat s}} \left({\hat s}^2 + {\hat t}^2 + {\hat s}{\hat t} \right)^2 \left( 48 c_1^2 + 11 c_2^2 + 40 c_1 c_2 \right) \, ,
\label{xsectionci}
\end{equation}
where $\sqrt{\hat s} = m_{\gamma \gamma}$, and ${\hat t}$ are the
usual Mandelstam invariants in $\gamma \gamma \to \gamma \gamma$ scattering. 
We see that this cross-section rises rapidly as a function of $m_{\gamma \gamma}$, conferring a competitive
advantage on a high-energy collider such as CLIC.

In the following Sections we describe the possible contributions to the effective Lagrangian coefficients $c_{1, 2}$
in (\ref{generaldim8}) from various types of new physics involving mass scales beyond the Standard Model.

\subsubsection{The Heisenberg-Euler Contribution from Massive Particles}
\label{sec:massiveloops}

 In the decoupling limit when $m_{\gamma \gamma}$ is much smaller than the masses of particles circulating in the loops, 
the resulting low-energy effective field theory is that of Heisenberg and Euler. The coefficients $c_{1,2}$ of the dimension-8 operators for electrically-charged particles of spin $S = 0, 1/2, 1$ and mass $M$ are given to leading order by~\cite{Fichetetal2014} 
\begin{equation}
c_i = \frac{\alpha_\text{EM}^2 Q_\text{eff}^4}{M^4} C^S_i \; , \; i = 1, 2 \, ,
\label{cCHE}
\end{equation}
where $Q_\text{eff}^4 \equiv \sum_i Q_i^4$ for particles $i$ of charges $Q_i$, and the spin-dependent coefficients
\begin{eqnarray}
C^0_1 &= & \frac{1}{288} \quad , \quad C^0_2 = \frac{1}{360} \, , \nonumber \\
C^\frac{1}{2}_1 &= & -\frac{1}{36} \quad , \quad C^\frac{1}{2}_2 = \frac{7}{90} \, ,\nonumber \\
C^1_1 &= & -\frac{5}{32} \quad , \quad C^1_2 = \frac{27}{40} \, .
\label{ciHE}
\end{eqnarray}
This approximation is sufficient to probe the sensitivity to particles that cannot be produced directly on-shell when $s \gg 4M^2$. Below this threshold the full contribution to the loop must be included. In the case of circulating quarks in the Standard Model amplitude, the leading-order QCD corrections have been calculated in~\cite{Bernetal}, and are included
in the {\tt SANC} code~\cite{SANC} that we use in this paper. This code also includes fully the effects of the $t$ and $W$ 
masses in their respective loop diagrams. In general, for such non-decoupled charged particles, one may write the unpolarized cross-section in the form:
\begin{equation}
\frac{d\sigma}{d\Omega} = \frac{\alpha_\text{EM}^4 Q_\text{eff}^8}{2\pi^2 s} \left( |M_{++++}|^2 + |M_{++--}|^2 + |M_{+-+-}|^2 + |M_{+--+}|^2 + 4|M_{+++-}|^2 \right) \, ,
\end{equation}
where expressions for the amplitudes for the various helicity combinations are given in Appendix A of Ref.~\cite{Fichetetal2014}.

\subsubsection{Born-Infeld Theory}

Born-Infeld theory~\cite{BornInfeld34} in $n$ dimensions postulates a non-polynomial Lagrangian given by 
\begin{equation}
{\cal L}_{\rm BI} \; = \; \beta^2 \Big(1 - \sqrt{-{\rm det} (\eta_{\mu\nu} + \frac{1}{\beta}F_{\mu\nu})}\, \Big) \, ,
\label{generalBI}
\end{equation}
where $\eta_{\mu\nu}$ is the Minkowski space-time metric, ${\det}$ denotes the appropriate determinant in
$n$ space-time dimensions, and $ \beta$ is an {\it a priori} unknown parameter that fixes
the maximum  field strength.
The general expression (\ref{generalBI}) may be written in the following form in the case of 4 space-time dimensions:
\begin{equation}
{\cal L}_{\rm BI} \; = \; \beta^2 \Big(1
 - \sqrt{1 + \frac{1}{2 \beta^2} F_{\mu \nu} F^{\mu \nu}
-\frac{1}{16  \beta^4} (F_{\mu \nu} \widetilde{F}^{\mu \nu})^2} \; \Big) \, ,
\label{LBI}
\end{equation}
where $\widetilde{F}^{\mu \nu}$ is the dual of the field strength tensor
${F_{\mu \nu}}$: $\widetilde{F}^{\mu \nu} \equiv \frac{1}{2} \epsilon^{\mu\nu\rho\sigma}\, F_{\rho\sigma}$.
In the four-dimensional case the parameter $\beta$ has the dimension of [Mass]$^2$ and can be written as $\beta \equiv M^2$,
where $M$ is a mass scale.

Expanding the Born-Infeld Lagrangian (\ref{LBI}) in 
inverse powers of $\beta$, we find operators of dimension 8 and higher in the effective field theory and thus 
make contact with (\ref{generaldim8}). For this purpose, it is convenient to
use the following representation of the 4-dimensional Born-Infeld theory (\ref{LBI})~\cite{tseytlin}:
\begin{eqnarray}\label{BI2}
{\cal L}_{\rm BI} &\simeq & 
- \beta^2 \, I_2 - \beta^2 \, I_4 \Big(1 + {\mathcal O}(F^2) \Big): \nonumber \\
 I_2 &=& \frac{1}{4\, \beta^2} F_{\mu\nu} \, F^{\mu\nu}~, \quad I_4 = -\frac{1}{8\, \beta^4} F_{\mu\nu} F^{\nu\rho} F_{\rho\lambda} F^{\lambda\mu} + \frac{1}{32\, \beta^4}  (F_{\mu\nu} \, F^{\mu\nu})^2 \, .
\end{eqnarray}
Expanding (\ref{LBI}) to fourth order in the electromagnetic field strength, we find the following
expressions for the coefficients $c_i$ appearing in (\ref{generaldim8}) in terms of $\beta$:
\begin{equation}
c_1 = - \frac{1}{32\, \beta^2} , \quad c_2 = \frac{1}{8\, \beta^2}~.
\label{cibeta}
\end{equation}
It follows from the structure (\ref{BI2}) that the ratio of these coefficients is $c_2/c_1=-4$ exactly, which is a characteristic prediction of Born-Infeld theory~\footnote{A similar relation between the coefficients may arise in a particular Heisenberg-Euler effective theory evaluated to this fixed order. However, no combination of massive particles with spin $\leq 1$ can be integrated out to reproduce the Born-Infeld relations exactly at higher orders~\cite{Hagiwara}.}.
Substituting the expressions (\ref{cibeta}) into the formula (\ref{xsectionci}), we recover the leading-order cross-section for 
unpolarised light-by-light scattering in Born-Infeld theory in the $\gamma \gamma$ centre-of-mass frame
as given by~\cite{Davilaetal,TurkRebhan,CT}:
\begin{equation}
\sigma_\text{BI}(\gamma\gamma \to \gamma\gamma) = \frac{1}{2}\int d\Omega \frac{d\sigma_\text{BI}}{d\Omega} = \frac{7}{1280\pi}\frac{m_{\gamma\gamma}^6}{\beta^4} \, ,
\label{BItotal}
\end{equation}
where $m_{\gamma\gamma}$ is the diphoton invariant mass and the differential cross-section is
\begin{equation}
\frac{d\sigma_\text{BI}}{d\Omega} =\frac{1}{4096\pi^2} \frac{m_{\gamma\gamma}^6}{\beta^4}\left(3 + \cos^2{\theta}\right)^2 \, .
\label{BIangle}
\end{equation}
We note again that this cross-section rises rapidly as a function of $m_{\gamma \gamma}$, conferring a competitive
advantage on a high-energy collider such as CLIC. This growth is scaled inversely by the 
dimensionful parameter $\beta = M^2$ that appears in the nonlinear Born-Infeld extension of 
QED defined by the Lagrangian (\ref{LBI}). If $\beta$ originates from a Born-Infeld theory of hypercharge then the 
corresponding mass scale is $M_Y = \cos{\theta_W} M$, where $\theta_W$ is the weak mixing angle.

The steep rise with $m_{\gamma \gamma}$ of the Born-Infeld cross-section (\ref{BItotal}) would be shared
by any other model that can be characterized via a dimension-8 interaction as in (\ref{generaldim8}) - see
(\ref{xsectionci}) - and is completely different from the energy dependence of the Heisenberg-Euler cross-section shown in the left panel
of Fig.~\ref{fig:sanc}. Likewise, the Born-Infeld angular distribution (\ref{BIangle}) is also shared with other combinations
of dimension-8 operator models - as also seen in (\ref{xsectionci}) - and is also completely different from the Standard Model prediction,
which is shown in the right panel of Fig.~\ref{fig:sanc}. Because of these differences, measurements of the two contributions 
to $\gamma \gamma$ scattering can be made in 
different regions of phase space, and it is a good approximation to neglect interference with Standard Model
scattering when we estimate later the $e^+ e^-$ collider sensitivity to the Born-Infeld scale.

\subsubsection{Magnetic Monopoles \label{sec:magmon}}

It was suggested in ~\cite{cho-maison} that monopole solutions could exist in a suitable regularisation of the
Standard Model. In the original formulation they were characterised by singular kinetic energy densities 
at the centre of the monopole, but these would be regularised by the non-linear
terms in a Born-Infeld extension of the hypercharge in the Standard Model~\cite{ArunasalamKobakhidze17}.
The finite-energy monopole solution is characterised by a mass 
\begin{equation}\label{fmm}
  M_{\cal M} = E_0 + E_1, 
 \end{equation}
where $E_0$ is the contribution associated with the
Born-Infeld U(1)$_{\rm Y}$ hypercharge, and $E_1$ is associated with the remainder of the Lagrangian.
The lower limit on $\beta$ obtained from light-by-light scattering in heavy-ion collisions
at the LHC~\cite{EMY2} indicates that $E_0 \gtrsim 6.2$~TeV.
It was estimated in~\cite{ArunasalamKobakhidze17} that 
\begin{equation}\label{ak}
E_0 \simeq 72.8 \, M_Y,  \quad M_Y = \cos{\theta_W} \sqrt{\beta}, 
\end{equation}
where $\beta$ is the QED Born-Infeld parameter. 
It was estimated in \cite{cho-maison}, based on a numerical solution for the magnetic monopole, that $E_1 \sim 4$~TeV, which
was also adopted by \cite{ArunasalamKobakhidze17}. However, 
the semi-analytic expression for the monopole solution presented in~\cite{sarkar} leads to an 
improved estimate of $E_1 \sim 7.6$~TeV.
We conclude that present data indicate that the Born-Infeld monopole mass is $\gtrsim 14$~TeV, beyond the reaches of
the LHC and the $e^+ e^-$ colliders considered here, but still potentially within reach of FCC-hh.

Loops of magnetic monopoles would contribute to light-by-light scattering, as shown in Fig.~\ref{fig:ggloop}.
We assume that their effects at energies $\ll M_{\cal M}$ can be calculated treating the 
monopoles as point-like.
The loop contributions of general point-like dyons of spin 1/2 have been considered in~\cite{russian},
who showed that the contribution of a point-like monopole of mass $M_{\cal M}$ may be written in the form 
\begin{equation}l
{\mathcal L}_{\rm EH}^{\rm EFT-\gamma-mono} = \frac{1}{36} (\frac{g}{\sqrt{4\pi}M_{\cal M}})^4 \Big(\frac{\widetilde \beta_+  + \widetilde \beta_-}{2} (F_{\mu\nu}\, F^{\mu\nu})^2 + \frac{\widetilde \beta_+  - \widetilde \beta_-}{2} (F_{\mu\nu}\, \widetilde F^{\mu\nu})^2 \Big) \, ,
\label{eq:Lmonopole}
\end{equation}
which is enhanced compared to the loops of electrically-charged particles considered earlier by the 
strong (quantized) magnetic coupling $g$ of the Dirac monopole~\cite{russian}: $g=\frac{2\pi n}{e}, n=\pm 1, \pm2, \dots $. 
We consider here the case of point-like monopoles of mass with spin 1/2 and
magnetic charge $n=\pm 1, \pm2, \dots $, for which the coefficients in (\ref{eq:Lmonopole}) take the values $\widetilde \beta_+ = 11/10$ and $\widetilde \beta_- = -3/10$~\cite{HeisenbergEuler36,russian}.
In this case, compared to the case considered in Section~\ref{sec:massiveloops} of spin-1/2 fermions 
of the same mass with effective
electric charge $Q_{\rm eff}$, the sensitivity is enhanced by a factor
\begin{equation}
\frac{g_M}{Q_{\rm eff}} \; = \; 2 \pi \frac{n}{Q_{\rm eff} e^2} \; = \; \frac{n}{2 \alpha_{\rm EM} Q_{\rm eff}}
\label{rescalemonopole}
\end{equation}
for a monopole of spin 1/2 and magnetic charge $n$.
More precisely, the Lagrangian \ref{eq:Lmonopole} with coefficients 
\begin{equation}
    \bar{c}_1 =  \frac{1}{36} (\frac{g}{\sqrt{4\pi}M_{\cal M}})^4 \frac{1}{2}\left(\widetilde \beta_+  + \widetilde \beta_-\right) \quad , \quad
    \bar{c}_2 =  \frac{1}{36} (\frac{g}{\sqrt{4\pi}M_{\cal M}})^4 \frac{1}{2}\left(\widetilde \beta_+  - \widetilde \beta_-\right) \, ,
\end{equation}
may be written in the basis of Eq.~\ref{generaldim8} using the relations $c_1 = \bar{c}_1 - 2 \bar{c}_2$, $c_2 = 4\bar{c}_2$. The enhancement in the cross-section \ref{xsectionci} for the point-like spin-1/2 monopoles relative to spin-1/2 fermions of the same mass is then given by 
\begin{equation}
    \frac{\sigma_\text{monopole}}{\sigma_{\text{fermion}}} = \left(\frac{1}{2}\frac{n}{\alpha_{\rm EM} Q_{\rm eff}} \right)^8 \, .
\end{equation}

The magnetic coupling $g_M$ is so large that it should in general
be treated non-perturbatively. However, the monopole-loop contribution to light-by-light scattering at low energies
 is suppressed by $1/M_{\cal M}^4$, and one expects in general that $M_{\cal M} \sim {\cal O}(m_W/\alpha)$ as is borne out by the
calculation (\ref{ak}). In this case, the monopole-loop contribution to low-energy light-by-light scattering is formally
comparable in magnitude to the perturbative contributions of electroweakly-interacting particles, and the naive one-loop calculation
may be taken as a good guide.

\section{Measurements of $\gamma \gamma \to \gamma \gamma$ at CLIC and FCC-ee}

\subsection{Experimental conditions}

In this study of $\gamma\gamma$ scattering, we consider the CLIC~\cite{CLICdp:2018cto} and FCC-ee~\cite{FCC:2018evy} collider options. The assumed centre-of-mass energy stages and corresponding integrated luminosities are summarised in Table~\ref{tab:luminosities}. For CLIC, the  integrated luminosities we use are those of the latest CLIC baseline scenario~\cite{Robson:2018zje}. Unpolarised lepton beams are assumed for both collider options throughout this paper.

Photon radiation off the initial-state leptons is described using the Equivalent Photon Approximation (EPA)~\cite{PeskinSchroeder}:
\begin{equation}
    f(z) = \frac{\alpha}{2\pi} \log \frac{s}{m^2} \left( \frac{1 + (1 - z)^2}{z} \right),
\end{equation}
where $z = E_\gamma / E_{\text beam}$ is the energy fraction of the radiated photon. In the following, the cut-off $z > 0.01$ is applied.
This is the only source of photons included for FCC-ee. 

The high bunch charge density at CLIC leads to strong photon radiation in the electromagnetic field of the other beam. This effect is referred to as Beamstrahlung in the following. The Beamstrahlung spectra which were prepared for the CLIC CDR with the \texttt{GUINEA-PIG} simulation code~\cite{Schulte:1999tx} are used here. The corresponding beam parameters are given in~\cite{Lebrun:2012hj}. The luminosities for the resulting collisions of Beamstrahlung photons with leptons of the other beam or of two Beamstrahlung photons are given in Table~\ref{tab:luminosities}. 

\begin{table}[h!]
\centering
\begin{tabular}{c | c | c | c | c} 
 Collider & $\sqrt{s}$ [GeV] & $\mathcal{L}_{e^+ e^-}$ [$fb^{-1}$] & $\mathcal{L}_{e^+ \gamma}$ = $\mathcal{L}_{\gamma e^-}$ [$fb^{-1}$] & $\mathcal{L}_{\gamma\gamma}$ [$fb^{-1}$] \\
 \hline
  & 350 & 1000 & 450 & 230 \\ 
 CLIC & 1400 & 2500 & 1875 & 1600 \\ 
  & 3000 & 5000 & 3950 & 3450 \\ 
 \hline
 FCC-ee & 365 & 1500 & - & - \\
 \hline
\end{tabular}
\caption{Assumed integrated luminosities for $e^+ e^-$, $e^\pm \gamma$ and $\gamma\gamma$ collisions at CLIC and FCC-ee, where the photons originate from the Beamstrahlung effect.}
\label{tab:luminosities}
\end{table}

\begin{figure}[h!]
\begin{center}
\includegraphics[scale=0.6]{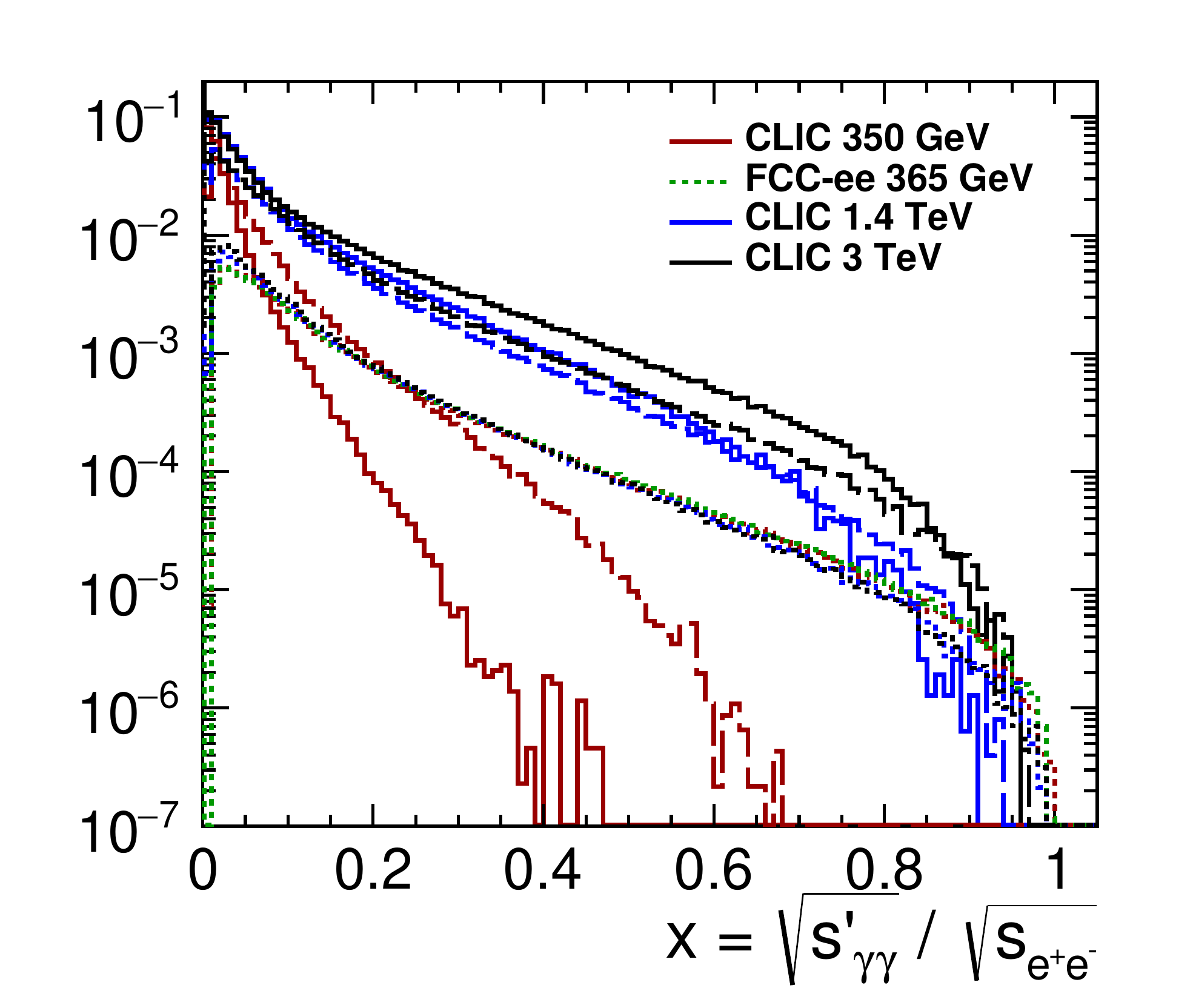}
\caption{\it The distributions of the $\gamma \gamma$ scattering centre-of-mass energy $\sqrt{s^\prime_{\gamma \gamma}}$
for FCC-ee at 365~GeV (green histogram) and for CLIC
at 350 GeV, 1.4 TeV and 3 TeV (red, blue and black histograms, respectively). In the latter cases,
the solid histogarms show the collisions of two Beamstrahlung photons, the dotted histograms the collisions of two EPA photons and the dashed histograms the collisions of one Beamstrahlung and one EPA photon. All histograms at a given energy stage are scaled to the same run time.
}
\label{fig:photon_spectra}
\end{center}
\end{figure}

In the case of CLIC, we have combined the contributions from collisions of two Beamstrahlung photons, two EPA photons or of one Beamstrahlung photon and one EPA photon were combined. 
The centre-of-mass energy distributions of the colliding photons at the different CLIC stages and FCC-ee operating at a
centre-of-mass energy of 365~GeV are compared in Fig.~\ref{fig:photon_spectra}. In the region of large $\sqrt{s^\prime_{\gamma \gamma}}$ 
that is most relevant for the new physics reach, collisions of EPA photons are dominant at CLIC operating
at 350~GeV while collisions involving Beamstrahlung photons are more important at the higher CLIC energies.

Interactions of virtual photons, which could be tagged by additional electrons in the final state, are not considered here.

\subsection{Signal simulation and background processes}

At each CLIC energy stage and for FCC-ee, one million beam-beam events were randomly generated for each considered initial-state combination ($\gamma_{\rm Beamstr}\gamma_{\rm Beamstr}$, $\gamma_{\rm Beamstr}\gamma_{\rm EPA}$, $\gamma_{\rm EPA}\gamma_{\rm Beamstr}$ and $\gamma_{\rm EPA}\gamma_{\rm EPA}$ for CLIC, only $\gamma_{\rm EPA}\gamma_{\rm EPA}$ for FCC-ee). These are used to construct $\gamma\gamma \to \gamma\gamma$ events, assuming a polar angle distribution $\propto (3 + \cos^2{\theta})^2$ for the dimension-8 operator contributions, where $\theta$ is the polar angle of a final-state photon in the centre-of-mass system of the hard interaction. The event rates are normalised using the minimum of  (\ref{xsectionci}) and $1/m_{\gamma\gamma}^2$, or (\ref{BItotal}) and $1/m_{\gamma\gamma}^2$ to suppress events at very large scales. In the Born-Infeld model the Effective Field Theory description may be expected to break down for $m_{\gamma\gamma} \simeq \sqrt{\beta}$, but in the Euler-Heisenberg case its validity depends on the magnitude of the electromagnetic coupling, as we discuss in more detail below.

The Standard Model contribution to $\gamma\gamma$ scattering described in Section~\ref{subsec:SM} represents 
an intrinsic background for new physics searches. Events are simulated according to the kinematics predicted by \texttt{SANC}. 
As for the dimension-8 operator contributions, beam-beam events are used as inputs to include 
the correlations between the two beams.

\begin{figure}[h!]
\begin{center}
\includegraphics[scale=0.6]{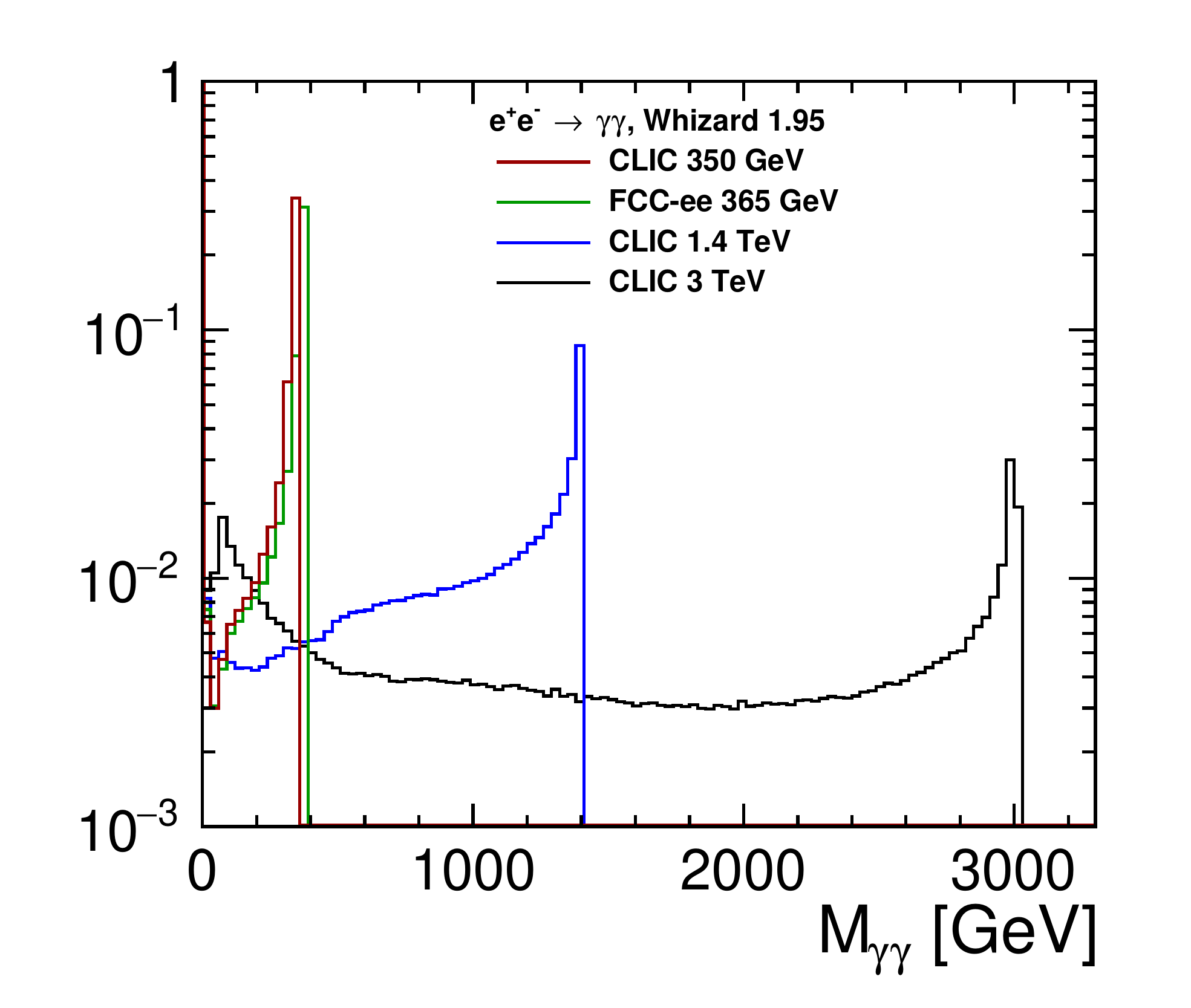}
\caption{\it The distributions of the $\gamma \gamma$ invariant mass in $e^+ e^- \to \gamma\gamma$ events 
at 350 GeV, 1.4 TeV and 3 TeV CLIC (red, blue and black histograms, respectively), and at 365 GeV FCC-ee (green histogram).
}
\label{fig:ee_background}
\end{center}
\end{figure}

A second crucial background process is $e^+ e^- \to \gamma\gamma$, which is simulated using the {\tt Whizard~1.95} package~\cite{Kilian:2007gr, Moretti:2001zz}. The obtained cross sections for $20^\circ < \theta_\gamma < 160^\circ$, where $\theta_\gamma$
is measured in the laboratory frame and $E_\gamma >$ 25 GeV, are 3300/3100/290/91 fb at 350 GeV CLIC/365 GeV FCC-ee/1.4 TeV CLIC/3 TeV CLIC.
Due to energy loss from the Beamstrahlung and ISR effects, the invariant mass of the final-state photon pair may be reduced significantly compared to the nominal centre-of-mass energy, as illustrated in Fig.~\ref{fig:ee_background}. While the distributions peak 
at the nominal electron-positron centre-of-mass energies, large tails extend towards lower values, particularly at the higher
CLIC centre-of-mass energies.

\subsection{Event selection}

To exclude kinematic regions dominated by the SM contribution to $\gamma\gamma$ scattering, 
the following requirements are imposed to the final-state photons in the laboratory frame:
\begin{itemize}
\item $20^\circ < \theta_\gamma < 160^\circ$ for each photon;
\item $E_\gamma > 25/25/200/400$~GeV for each photon at 350 GeV CLIC/365 GeV FCC-ee/1.4 TeV CLIC/3 TeV CLIC.
\end{itemize}
In addition, the $e^+ e^- \to \gamma\gamma$ background was reduced using a cut on the invariant mass of the 
two final-state photons:
\begin{itemize}
\item $M_{\gamma\gamma} < 310/330/1150/2650$~GeV for each photon at 350 GeV CLIC/365 GeV FCC-ee/1.4 TeV CLIC/3 TeV CLIC.
\end{itemize}
No detector effects are included in the present study, and it is assumed that all photons are reconstructed. 
In full simulation studies of realistic CLIC~\cite{Arominski:2018uuz} and FCC-ee~\cite{Bacchetta:2019fmz} detector concepts, 
photon efficiencies well above 95\% are found for photons passing the event selection cuts described above.
The expected photon energy resolutions of the CLIC and FCC-ee detector concepts of better than 
$\sigma(E)/E = 20\%/\sqrt{E} \oplus 1\%$ are not expected to have a sizeable impact on the measurement of 
light-by-light scattering using the cut and count approach described above.

In the kinematic region defined by the cuts described above, the 
SM cross-section for light-by-light scattering can be measured with a statistical 
precision of 29\%/56\%/15\%/5\% at 350 GeV CLIC/365 GeV FCC-ee/1.4 TeV CLIC/3 TeV CLIC 
in the absence of new physics contributions. CLIC at 350 GeV achieves greater precision than the higher-luminosity FCC-ee at 365 GeV because of the steep increase in the SM cross section as $m_{\gamma\gamma}$ decreases (see Fig. 2). In the low $m_{\gamma\gamma}$ region, Beamstrahlung is relevant even at 350 GeV CLIC (see Fig. 3), improving the precision of 350 GeV CLIC compared to 365 GeV FCC-ee.
A more precise determination of the SM cross-section 
could be achieved using an optimised selection for this purpose, but this is beyond the scope of the present analysis.

\section{Sensitivities to New Physics in $\gamma \gamma$ Scattering}

\subsection{Sensitivity to Generic Dimension-8 Operators}

\begin{figure}[h!]
\begin{center}
\vspace{1cm}
\includegraphics[scale=0.6]{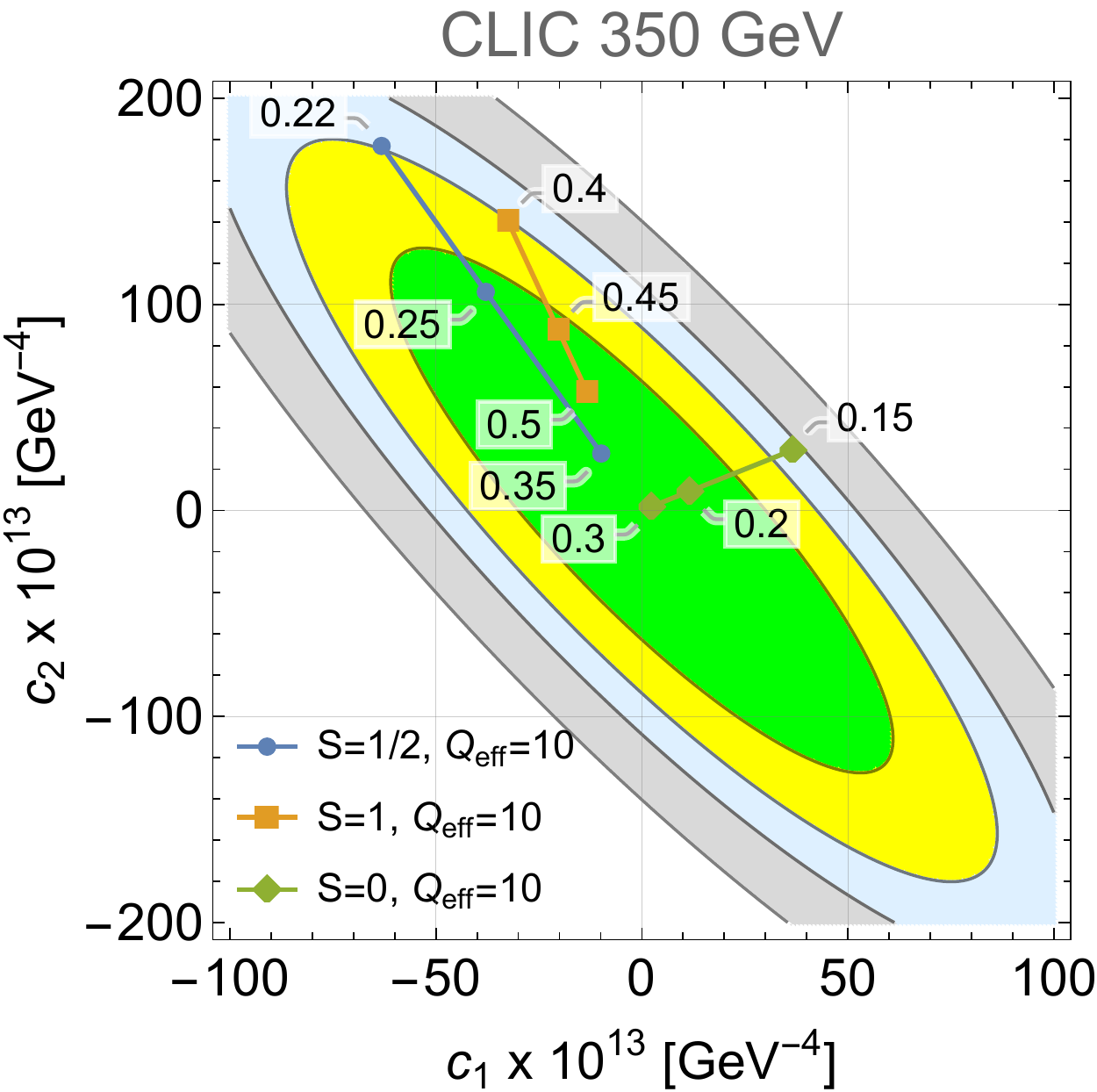}
\includegraphics[scale=0.6]{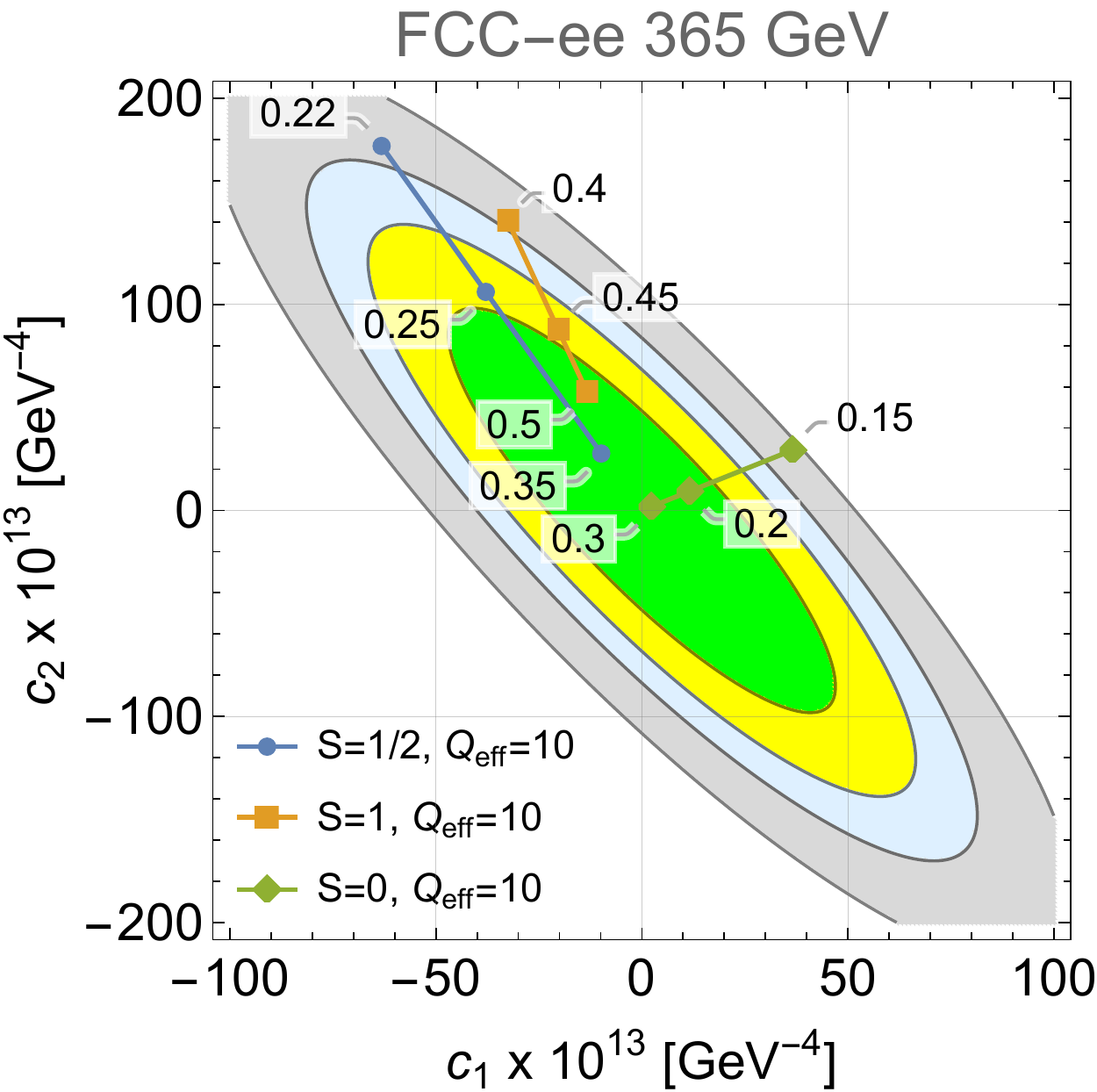}\\
\vspace{0.25cm}
\includegraphics[scale=0.6]{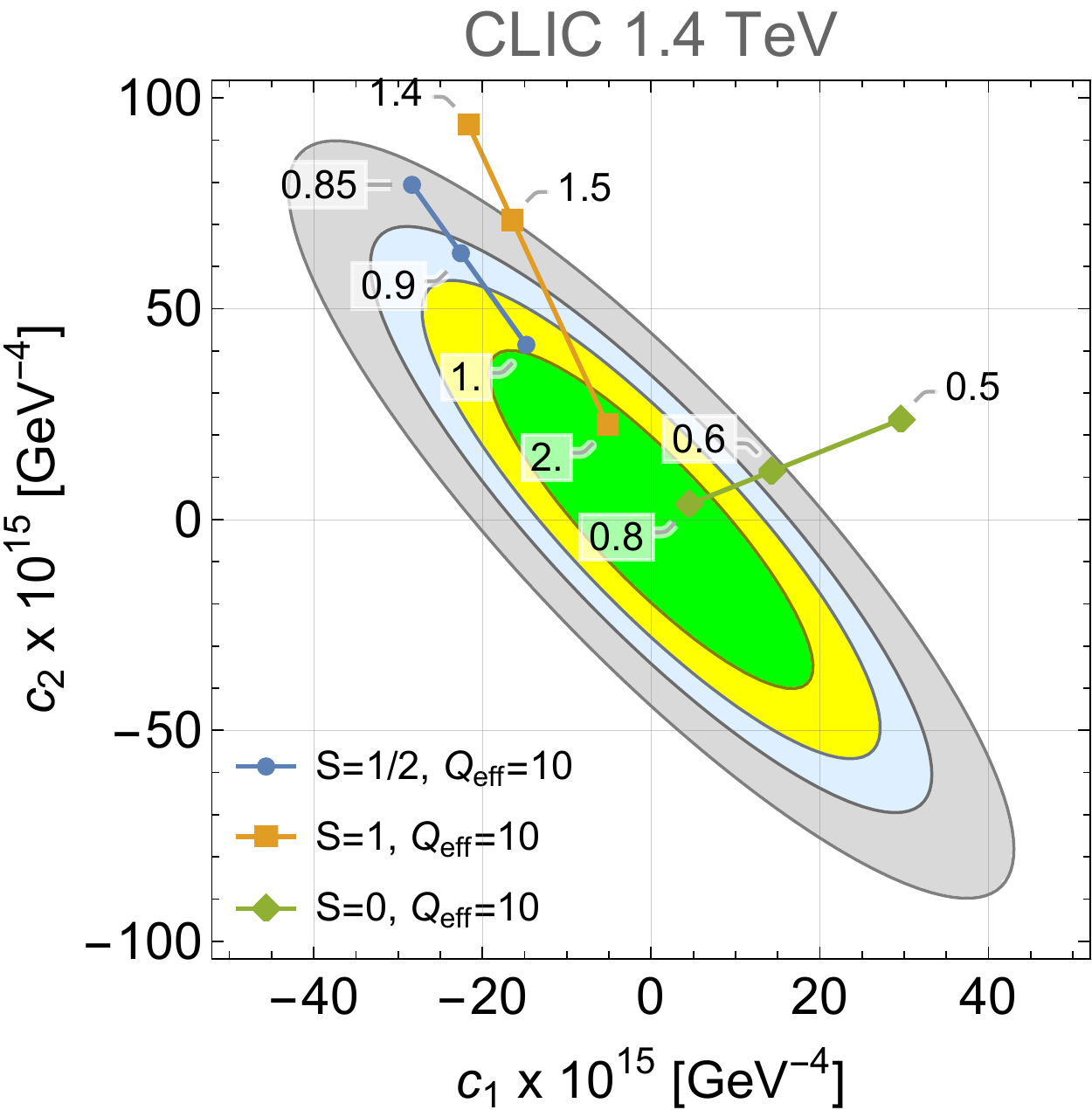}
\includegraphics[scale=0.6]{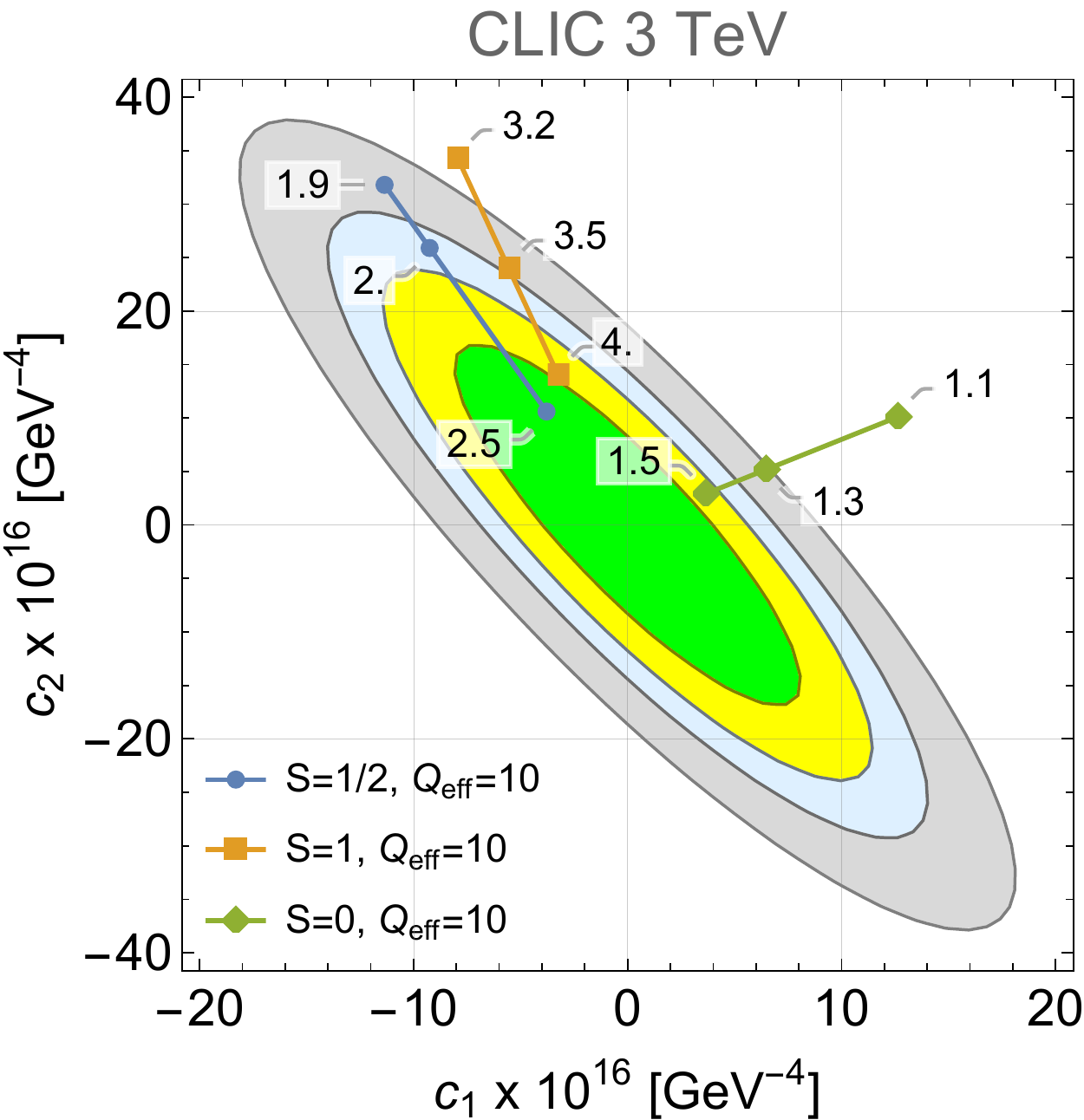}
\caption{\it Contours in the $(c_1, c_2)$ plane of 1-, 2-, 3-, and 5-$\sigma$ significance sensitivities
of possible measurements with FCC-ee and CLIC at 350~GeV are
shown in the upper panels, and those of possible CLIC measurements at centre-of-mass energies
$\sqrt{s} = 1.4$ and 3~TeV are shown in the lower panels. The potential 5-$\sigma$ discovery region is outside the grey contour,
and the potential 95\% exclusion reach is outside the yellow contour. Also shown are 
the lines corresponding to the Heisenberg-Euler~\cite{HeisenbergEuler36} contributions of heavy spin 0, -1/2 and -1 
beyond-the-Standard-Model particles in blue, orange, and green, respectively, with effective charge $Q_\text{eff} = 10$, a value that has been chosen for consistency with the validity of the relevant EFT. 
The dots are labelled with the masses of the particles in units of TeV. }
\label{fig:heisenbergeulercontourplot}
\end{center}
\end{figure}

The prospective sensitivities of $e^+ e^-$ collider measurements to the possible dimension-8 operator coefficients (\ref{generaldim8})
in the $(c_1, c_2)$ plane are shown in Fig.~\ref{fig:heisenbergeulercontourplot}. Those of FCC-ee and CLIC at 350~GeV are
shown in the upper panels, and those of CLIC measurements at centre-of-mass energies
$\sqrt{s} = 1.4$ and 3~TeV are shown in the lower panels.
We display contours of 1-, 2-, 3-, and 5-$\sigma$ statistical significance, 
calculated as $S/\sqrt{B}$ where $S$ is the number of dimension-8 signal events and
$B$ is the number of background events. The 5-$\sigma$ contour represents
the potential discovery sensitivity in the presence of a signal, and the 2-$\sigma$ contour represents the potential 
95\% CL exclusion contour in the absence of any signal. We see that typical 95\% CL sensitivities to the dimension-8 operator
coefficients are $c_{1,2} \lesssim {\cal O}(10^{-11})$~GeV$^{-4}$ for FCC-ee and CLIC at 350~GeV, which
improve to $c_{1,2} \gtrsim {\cal O}(4 \times 10^{-14})$~GeV$^{-4}$ for CLIC at 1.4 TeV and $c_{1,2} \gtrsim {\cal O}(2 \times 10^{-15})$~GeV$^{-4}$ at 3~TeV. The latter
correspond to sensitivities to new physics at mass scales $\sim 2-5$~TeV. The 1.4 TeV and 3 TeV stages of CLIC would therefore improve over the bounds set 
recently by the TOTEM and CMS Collaborations at the LHC~\cite{TOTEM:2021kin}. However, the FCC-ee and the 350 GeV stage of CLIC are not as sensitive as the LHC. Projections for HL-LHC have been estimated in~\cite{Fichetetal2014}.

In the following Sections we discuss the interpretations of the FCC-ee and CLIC sensitivities in the
theoretical frameworks discussed earlier.

\newpage
\subsection{Sensitivities to Heisenberg-Euler Contributions} 
\label{sec:HE}

The lines in Fig.~\ref{fig:heisenbergeulercontourplot} represent the possible contributions
to the Heisenberg-Euler Lagrangian from heavy particles of spin 0, 1/2 and 1 and effective charge $Q_{\text{eff}}=10$, 
defined below Eq.~(\ref{cCHE}), that may arise in scenarios for new physics beyond the Standard Model.\footnote{This illustrative value is chosen to ensure consistency with the Effective Field Theory (EFT) approximation. Extending the analysis to smaller $Q_{eff}$ could either correspond to a simple rescaling or involve a more involved analysis beyond the EFT approximation, depending on the sensitivity. We leave such an investigation to future studies.} 
A large effective charge enables sensitivity to the highest scales, since it compensates for the 
loop factor in the heavy particle contributions.  

The dots along the lines represent
the values of $(c_1, c_2)$ expected for the heavy particle masses indicated (in TeV) by the labels.
The directions in the $(c_1, c_2)$ plane differ for the different spins: unfortunately they cannot be
distinguished experimentally, since the kinematical distributions for the different spins are identical
in the decoupling limit, as can be seen in (\ref{xsectionci})~\footnote{We note that the formulae for the $c_i$
given in (\ref{ciHE}) are valid in the large-mass limit, and one should take sub-asymptotic corrections
into account when making more precise estimates of the sensitivities to the masses of new particles
of different types.}.

In the case of spin 0, we see that the 5-$\sigma$ discovery sensitivity for $\sqrt{s} = 1.4$~TeV
is to a mass $m_0 \lesssim 0.6$~TeV, extending somewhat beyond the reach for direct pair-production. On the other hand,
at $\sqrt{s} = 3$~TeV the CLIC discovery reach for spin-0 particles would be to $m_0 \simeq 1.3$~TeV,
less than the reach for pair-production. In the cases of spins 1/2 and 1, for $\sqrt{s} = 1.4$~TeV the 5-$\sigma$ discovery reaches
again extend beyond the pair-production threshold to a mass $m_{1/2} \simeq 0.85$~TeV for spin 1/2 and $m_1 \simeq 1.5$~TeV for spin 1, and for $\sqrt{s} = 3$~TeV the
direct discovery reach for spin 1/2 is $m_{1/2} \simeq 1.8$~TeV whereas for spin 1 it is $m_1 \simeq 3.4$~TeV. For spin 1 the masses are sufficiently
far above the centre of mass energy that the EFT approach should be a good approximation. In the spin-0 and -1/2 cases the sensitivities are greater for larger effective electromagnetic charges. However, we note that most of the sensitivity comes from lower diphoton 
centre-of-mass energies where there are higher statistics, so the EFT approach may provide a good first approximation even for smaller
electromagnetic charges.

\subsection{Constraint on Born-Infeld Theory \label{sec:BI}}

\begin{figure}[h!]
\begin{center}
\includegraphics[scale=0.6]{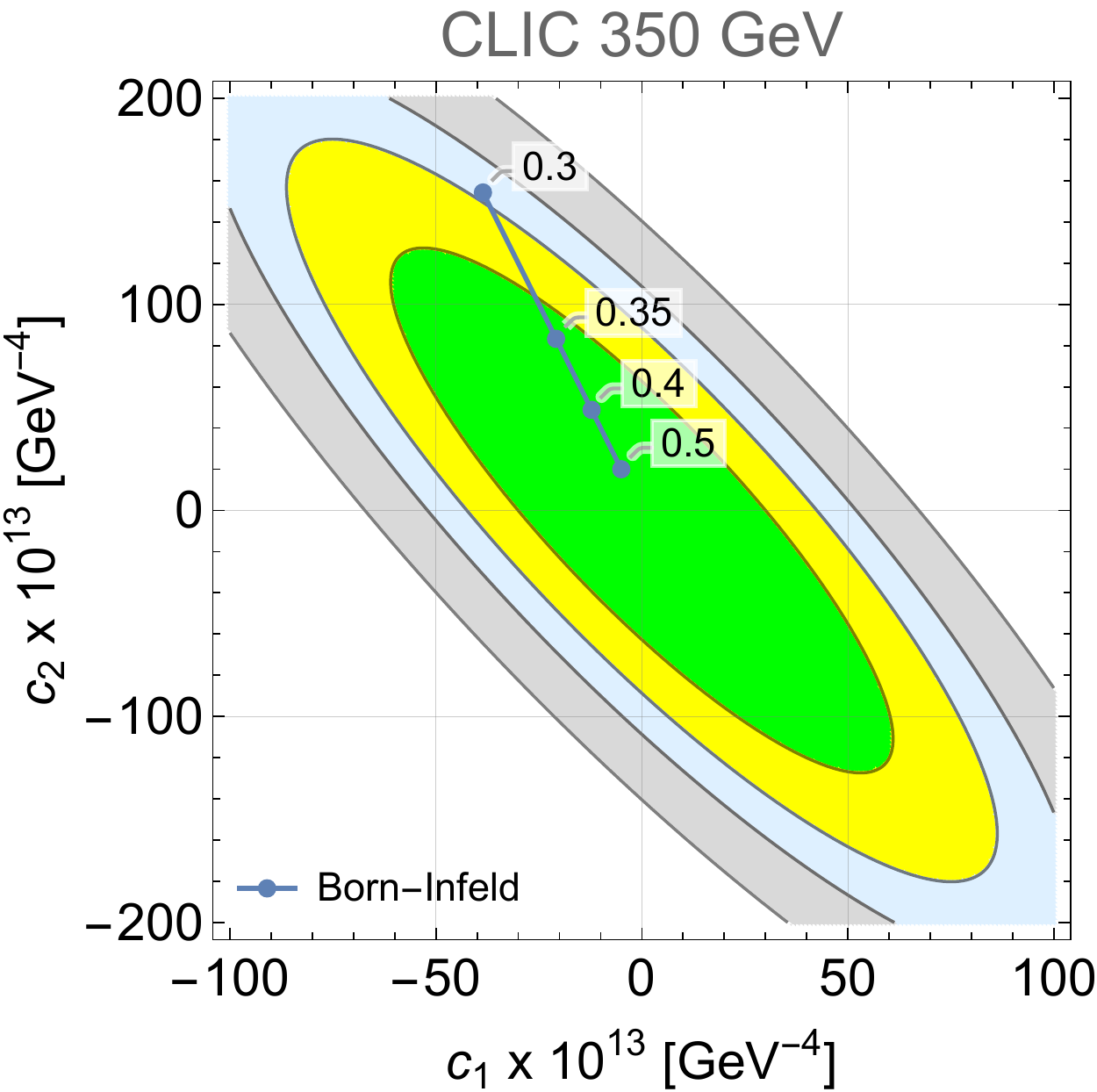}
\includegraphics[scale=0.6]{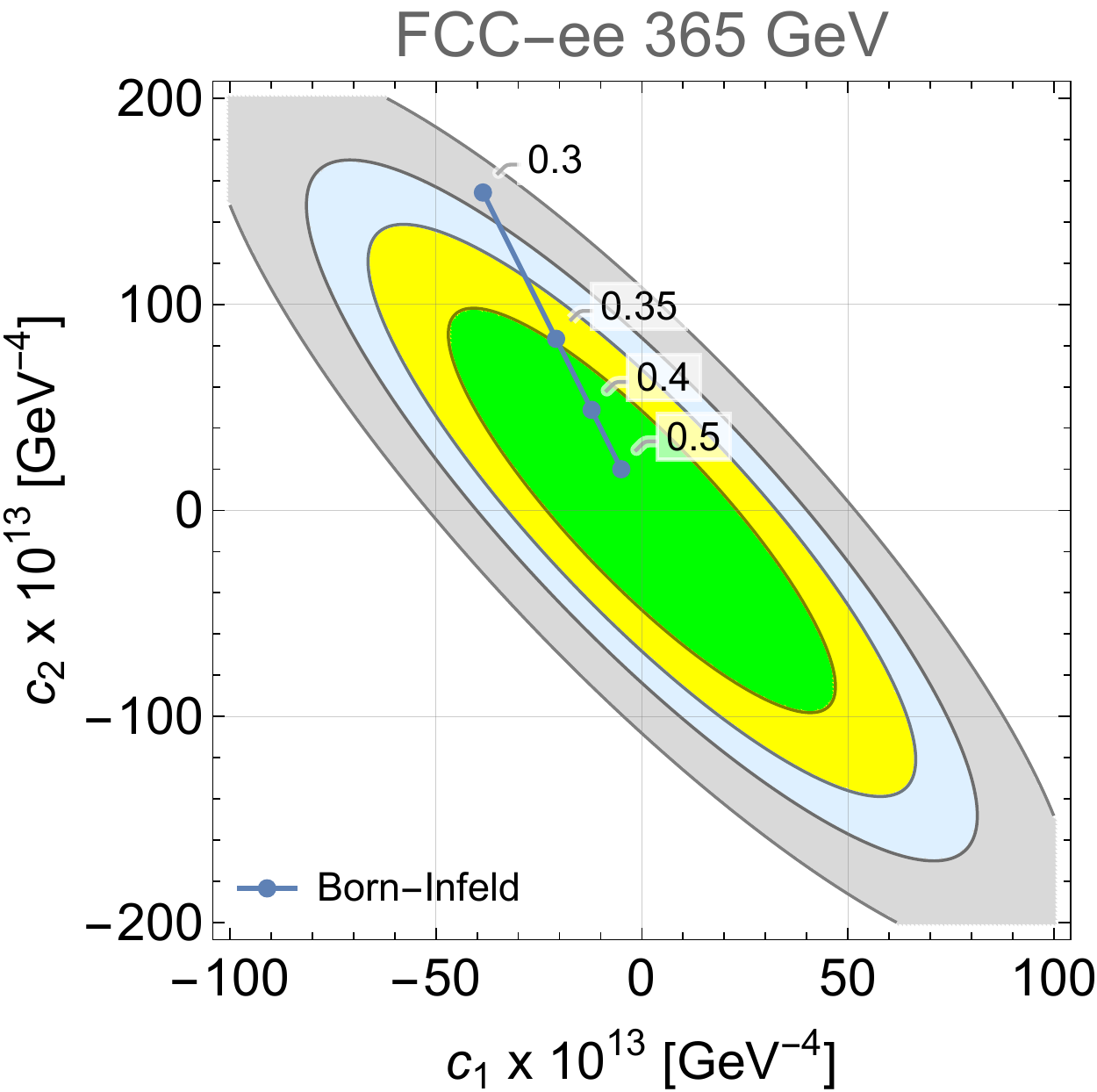}\\
\vspace{0.25cm}
\includegraphics[scale=0.6]{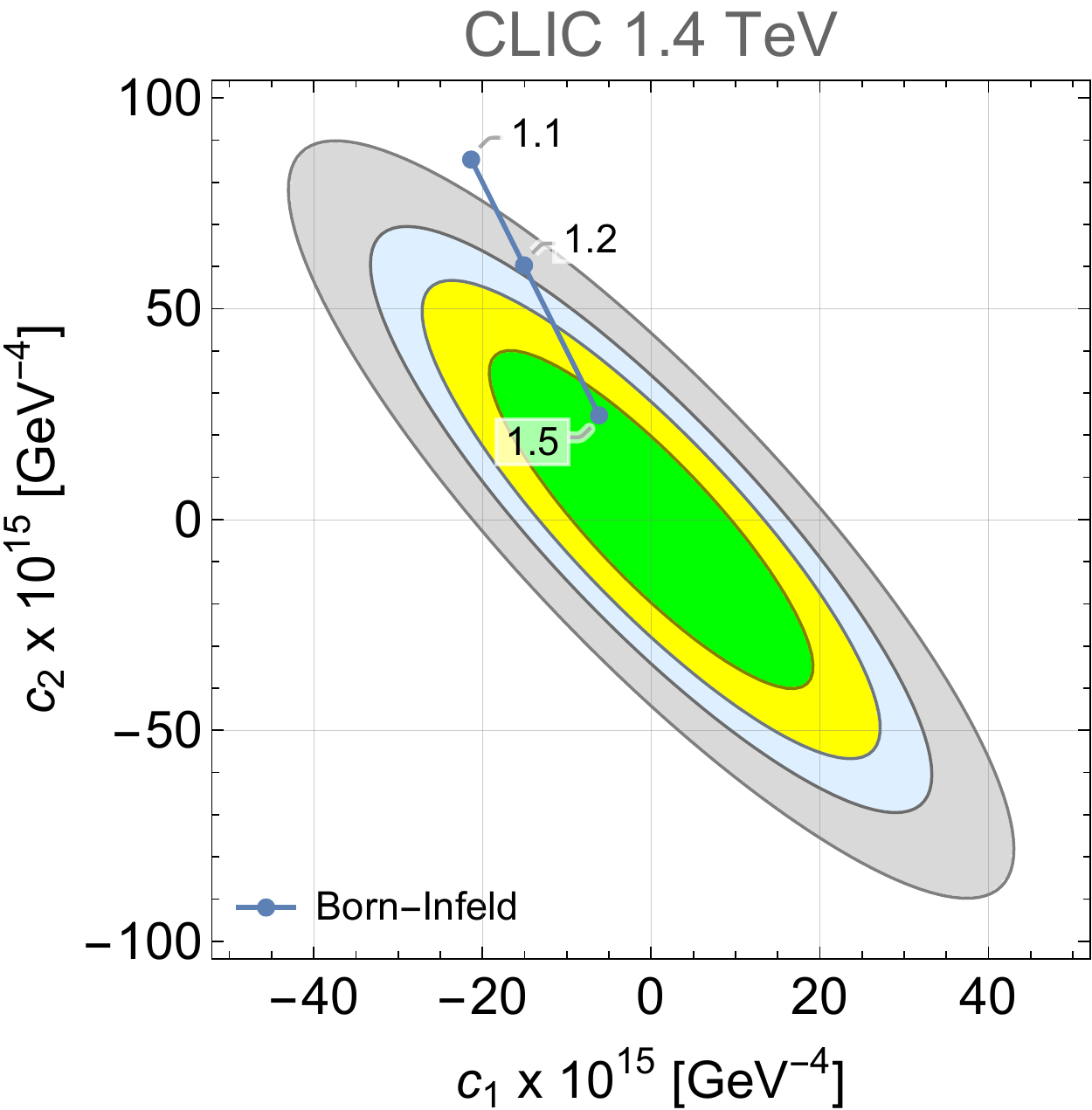}
\includegraphics[scale=0.6]{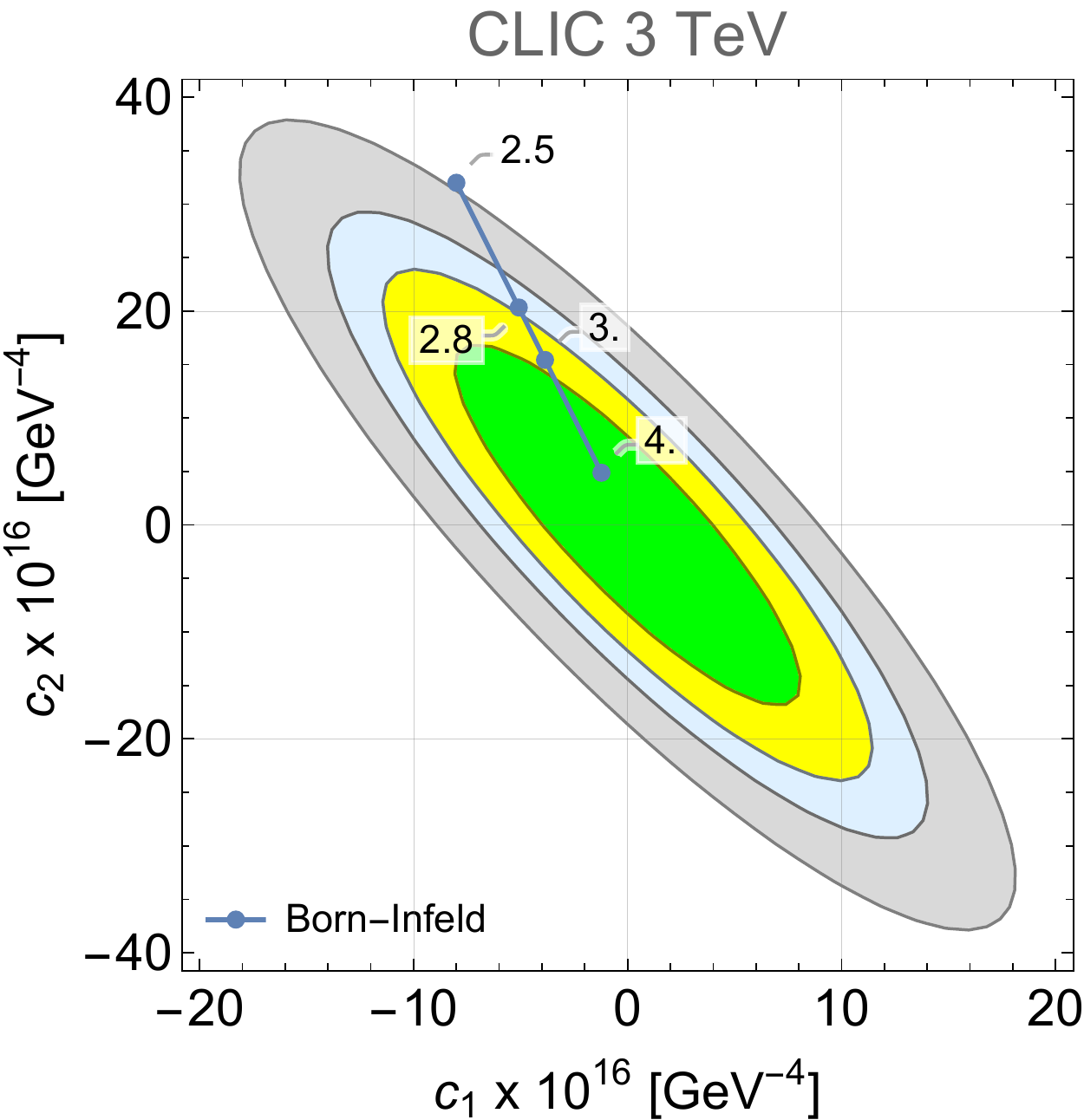}
\caption{\it Contours in the $(c_1, c_2)$ plane of 1-, 2-, 3-, and 5-$\sigma$ significance sensitivities
of possible measurements with FCC-ee and CLIC at 350~GeV are
shown in the upper panels, and those of possible CLIC measurements at centre-of-mass energies
$\sqrt{s} = 1.4$ and 3~TeV are shown in the lower panels. The potential 5-$\sigma$ discovery region is outside the grey contour,
and the potential 95\% exclusion reach is outside the yellow contour. Also shown are 
the lines corresponding to the Born-Infeld model~\cite{BornInfeld34}, for which $c_2/c_1 = - 4$.
The dots are labelled with the corresponding values of the Born-Infeld scale in units of TeV. }
\label{fig:borninfeldcontourplot}
\end{center}
\end{figure}

\begin{figure}[h!]
\begin{center}
\includegraphics[scale=0.35]{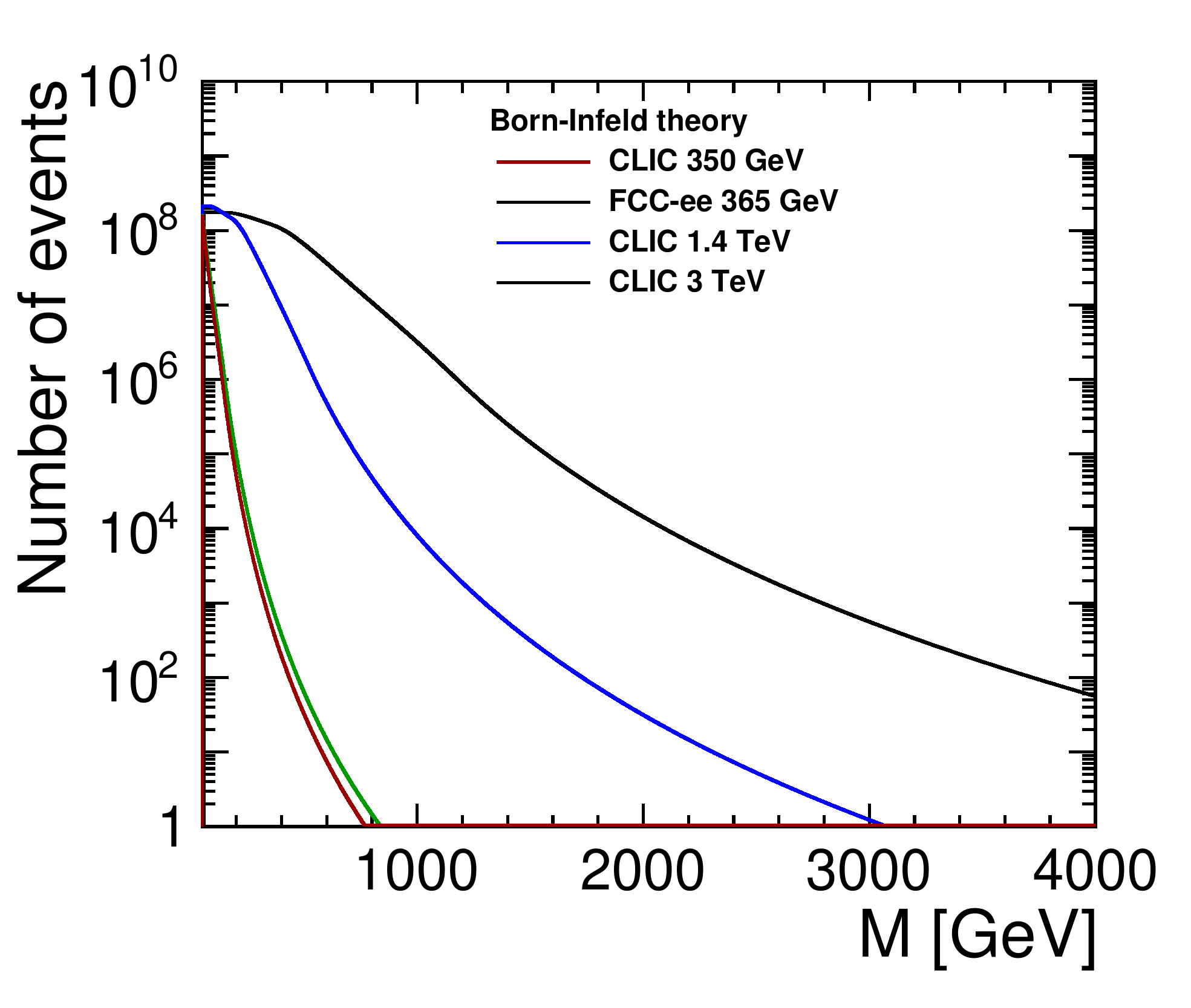} 
\includegraphics[scale=0.35]{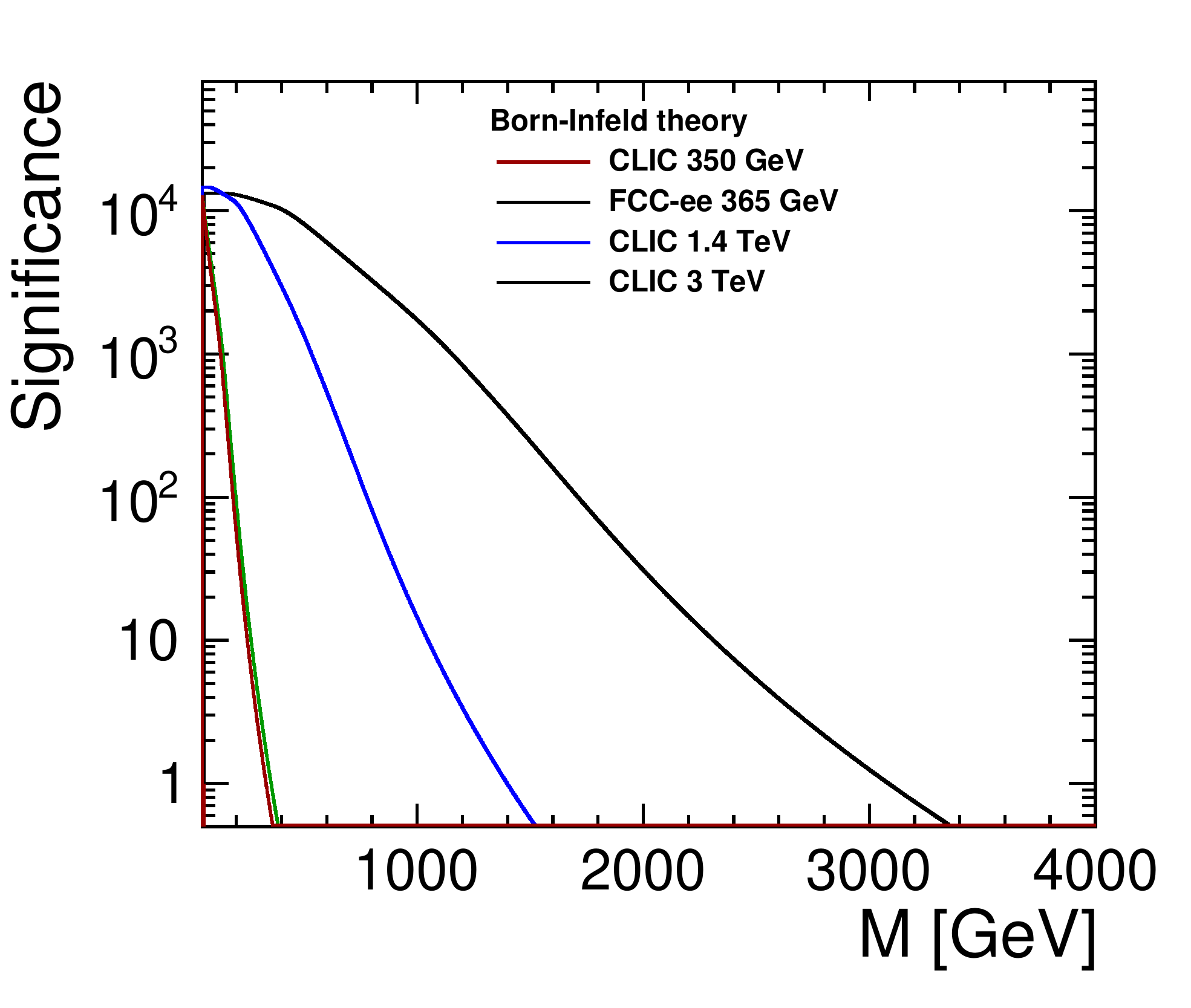}
\caption{\it Left panel: The number of $\gamma \gamma \to \gamma \gamma$ events in Born-Infeld theory~\cite{BornInfeld34} 
as a function of the Born-Infeld scale M at 350 GeV CLIC, FCC-ee, 1.4 TeV CLIC and 3 TeV CLIC 
(red, green, blue and black histograms, respectively). Right panel: The statistical significances 
of excesses of events compared to the SM background as a function of the Born-Infeld scale 
M at 350 GeV CLIC, FCC-ee, 1.4 TeV CLIC and 3 TeV CLIC 
(red, green, blue and black histograms, respectively).}
\label{fig:sensitivity_born_infeld}
\end{center}
\end{figure}

We discuss now the sensitivity of CLIC and FCC-ee measurements of $\gamma \gamma$ scattering to the
mass parameter $M = \sqrt{\beta}$ that characterizes the Born-Infeld Lagrangian (\ref{LBI}). 
Fig.~\ref{fig:borninfeldcontourplot} displays the same contours as in the previous section 
with the trajectory corresponding to Born-Infeld theory~\cite{BornInfeld34} at various scales, and
Fig.~\ref{fig:sensitivity_born_infeld} shows the number of events produced 
as a function of the Born-Infeld mass parameter $M$. The right panel shows the statistical
significance---calculated as $S/\sqrt{B}$ where $S$ is the number of Born-Infeld events and
$B$ is the number of background events---of a prospective Born-Infeld 
signal excess above the Standard Model background. 
A significance $S \ge 5$ corresponds to a $\ge 5$-$\sigma$ discovery 
of new physics, and $S = 2$ corresponds to the 95\% CL lower limit on $M$ in the absence
of any signal. On the basis of Fig.~\ref{fig:sensitivity_born_infeld} we estimate for 3 TeV CLIC:
\begin{eqnarray}
5\sigma \, {\rm discovery~range}: & 2.5~{\rm TeV}, & 95\% \, {\rm CL~lower~limit}: \; 2.8~{\rm TeV} \, .
\label{3000reach}
\end{eqnarray}
As seen in Fig.~\ref{fig:sensitivity_born_infeld}, we have made similar estimates of the CLIC and FCC-ee sensitivities
to the Born-Infeld scale from running at the lower energies of 350, 365 and 1400 GeV:
\begin{eqnarray}
{\rm CLIC~350~GeV}: 5\sigma \, {\rm discovery~range}: & 280~{\rm GeV}, \, 95\% \, {\rm CL~lower~limit}: \; 310 ~{\rm GeV} \, , \nonumber \\
{\rm FCC-ee~365~GeV}: 5\sigma \, {\rm discovery~range}: & 290~{\rm GeV}, \, 95\% \, {\rm CL~lower~limit}: \; 330 ~{\rm GeV} \, , \nonumber \\
{\rm CLIC~1.4~TeV}: 5\sigma \, {\rm discovery~range}: & 1.14~{\rm TeV}, \, 95\% \, {\rm CL~lower~limit}: \; 1.31 ~{\rm TeV} \, . \nonumber \\
\label{otherreaches}
\end{eqnarray}
We see that the CLIC sensitivities at 3~TeV and 1.4~TeV in the centre-of mass extend to Born-Infeld scales above
1~TeV, into the range of relevance for low-scale brane scenarios.

\subsection{Constraints on Monopole Masses}

\subsubsection{Sensitivity from Monopole Loops}

In the case of CLIC at 3~TeV, we found in Section~\ref{sec:HE}
a 95\% CL sensitivity to spin-1/2 fermions with $Q_{\rm eff} =10$ and mass 2.1~TeV. Using the rescaling relation (\ref{rescalemonopole}), this corresponds to a mass
\begin{equation}
M_{\cal M} \; \sim \; 14 \, |n|~{\rm TeV} \, 
\label{Loopmasslimit}
\end{equation}
for a monopole of spin 1/2 and magnetic charge $n =\pm 1, \pm 2, ...$.
This is beyond the reach of the LHC, though
probably within the reach of FCC-hh~\cite{FCC-hh} for $|n| = 1$ or 2.

\subsubsection{Indirect Sensitivity within Born-Infeld Theory}

The sensitivities to the Born-Infeld theory~\cite{BornInfeld34} presented in Section~\ref{sec:BI} 
can be translated into the corresponding monopole masses, as discussed in~\cite{EMY2}.
For example, a Born-Infeld scale of 4~TeV, which is intermediate between the 5-$\sigma$ discovery
sensitivity and the 95\% CL exclusion sensitivity of CLIC at 3~TeV estimated in
Section~\ref{sec:BI}, would correspond to a monopole mass
\begin{equation}
M_{\cal M} \; \gtrsim \; (\xi  + 72.8 \times \cos \theta_W) \times 4~{\rm TeV},
\label{BImasslimit}
\end{equation}
where $\xi \sim 1$ for the numerical solution of \cite{cho-maison}, yielding $M_{\cal M} \; \sim \; 260~{\rm TeV} \, $, 
and $\xi \sim 7.6/4 = 1.9$ for the semi-analytic solution of \cite{sarkar}, yielding  $M_{\cal M} \; \sim \; 263~{\rm TeV}$, 
beyond the reach of any collider currently contemplated~\footnote{Above, we used the represntative value  $\sin^2\theta_W=0.23$;  
more accurately, we should use the value of $\sin \theta_W$ at the CLIC energy scale, which we estimated to be $\sin^2\theta_W=0.25$, 
however doing so the result changes only by about 1\% .}. This is also the case for
Born-Infeld monopoles when we consider the prospective CLIC sensitivity at 1.4~TeV, 
which we estimate to be to a Born-Infeld scale $\sim 1.3$~TeV, corresponding to a monopole mass $\sim 90$~TeV.
However, the prospective CLIC sensitivities at 350~GeV is to 
a Born-Infeld scale $\sim 300$~GeV, corresponding to a monopole mass $\sim 26$~TeV, which
might be accessible to a future high-energy proton-proton collider such as FCC-hh~\cite{FCC-hh}.

The constraint (\ref{BImasslimit}) is much stronger than that derived previously (\ref{Loopmasslimit}) for a point-like spin-1/2 monopole.
However, (\ref{BImasslimit}) only applies to a Born-Infeld monopole, whereas (\ref{Loopmasslimit}) applies to a wider class
of monopoles. Moreover, the validity of the point-like approximation for the monopole solutions discussed 
in~\cite{cho-maison,ArunasalamKobakhidze17} may be questioned.
Hence the two limits (\ref{BImasslimit}) and (\ref{Loopmasslimit}) should be regarded as complementary.
They demonstrate the versatility of the prospective sensitivity of CLIC to new effects in light-by-light scattering.

\section{Conclusions}

We have explored in this paper the sensitivities of prospective FCC-ee and CLIC measurements of light-by-light
scattering to possible new physics. This could manifest itself either via loop diagram contributions to the effective QED
Lagrangian as calculated by Heisenberg and Euler~\cite{HeisenbergEuler36}, or via dimension-8 effective
operators as appear in Born-Infeld theory~\cite{BornInfeld34}. The contemporary interest of the latter possibility has
been reinforced by the realization that Born-Infeld theory emerges naturally in string theory~\cite{tseytlin},
most specifically in the D-brane framework~\cite{Bachas95}.

While we find that FCC-ee will not be competitive with the LHC~\cite{TOTEM:2021kin}, 
CLIC has two advantages for the measurement of light-by-light scattering, as compared to FCC-ee. 
One is its higher centre-of-mass energy, which is advantageous for measuring the strongly
energy-dependent effects of both the Heisenberg-Euler Lagrangian and generic dimension-8 operators
such as appear in Born-Infeld theory. A second advantage is that the CLIC photon spectrum has an important
Beamstrahlung contribution in addition to the EPA contribution that is present also for lower-energy
circular $e^+e^-$ colliders.

We have found that CLIC measurements could be sensitive to generic dimension-8 operators with coefficients up to
${\cal O}(2 \times 10^{-15})$~GeV$^{-4}$ at the 95\% CL, corresponding to a new physics mass scale $\sim 5$~TeV. Interpreting
this sensitivity in the framework of the Heisenberg-Euler Lagrangian, we find that light-by-light
scattering at CLIC could be sensitive to new {large charge/multiplicity} spin-1 bosons with masses beyond its kinematic
reach, and similarly to spin-1/2 bosons with masses comparable to its beam energy. Interpreting
the CLIC sensitivity within Born-Infeld theory confirms that light-by-light measurements at a centre-of-mass energy of 3 TeV
would provide 5-$\sigma$ discovery sensitivity to a Born-Infeld scale $\sqrt{\beta} = M = 2.5$~TeV.
This discovery sensitivity is particularly interesting for some D-brane scenarios that envisage mass scales
in the TeV range.

An interesting application of this analysis is to constrain the possible mass of a magnetic monopole.
Assuming that its low-energy effects are the same for a point-like particle, the
rescaling (\ref{rescalemonopole}) of the sensitivity for electrically-charged particles corresponds to a
sensitivity to loops of spin-1/2 monopoles of charge $|n|$ of mass $M_{\cal M} \sim 14 |n|$~TeV.
However, in the specific framework of Born-Infeld theory, the CLIC light-by-light sensitivity at 1.4 (3) TeV
corresponds to a Born-Infeld monopole mass of 90 (260)~GeV.

This analysis reinforces the key message of~\cite{CLIC6}, namely that the higher centre-of-mass energy of
CLIC gives it a particular competitive advantage in the search for new physics manifested by 
higher-dimensional operators, compared to lower-energy $e^+ e^-$ colliders. This was first illustrated in~\cite{CLIC6}
in the case of dimension-6 operator coefficients, and here we have extended this observation to dimension-8
operators, in which case CLIC benefits also from its enhanced photon spectrum due to Beamstrahlung. 
In general, the high energy of CLIC gives it a long lever arm for these indirect searches for new physics~\cite{BB,BB2,BB3,Jik},
as well as for direct pair-production searches.

\section*{Acknowledgements}

The work of JE and NEM was supported partly by STFC Grants ST/L000258/1 and ST/T000759/1. The work of JE was 
also supported partly by the Estonian Research Council via a Mobilitas Pluss grant.
TY is supported by a Branco Weiss Society in Science Fellowship and partially supported by STFC consolidated grant ST/P000681/1. JE and NEM also acknowledge participation in the COST Association Action CA18108 ``{\it Quantum Gravity Phenomenology in the Multimessenger Approach (QG-MM)}''.

\end{document}